\documentclass[aps,prx,twocolumn,superscriptaddress,longbibliography]{revtex4}
\usepackage{graphicx}
\usepackage{latexsym}
\usepackage{amssymb}
\usepackage{amsmath}
\usepackage{amsfonts}
\usepackage{upgreek}
\usepackage{bm}
\usepackage{multirow}
\usepackage{enumitem}
\usepackage{color}
\usepackage{hyperref}
\hypersetup{
colorlinks = true,
linkcolor = [rgb]{0.70,0.13,0.13},
citecolor = [rgb]{0.13,0.55,0.13},
urlcolor = [rgb]{0.25, 0.41, 0.88}}

\newcommand{\dd}{\mathrm{d}}
\newcommand{\ii}{\mathrm{i}}

\newcommand{\U}{\mathrm{U}}
\newcommand{\SU}{\mathrm{SU}}
\renewcommand{\O}{\mathrm{O}}
\newcommand{\SO}{\mathrm{SO}}
\newcommand{\Sp}{\mathrm{Sp}}
\newcommand{\dsZ}{\mathbb{Z}}

\newcommand{\scL}{\mathcal{L}}
\newcommand{\scM}{\mathcal{M}}
\newcommand{\scO}{\mathcal{O}}

\newcommand{\scK}{\mathcal{K}}
\newcommand{\scT}{\mathcal{T}}

\newcommand{\sgn}{\operatorname{sgn}}
\renewcommand{\Re}{\operatorname{Re}}
\renewcommand{\Im}{\operatorname{Im}}
\newcommand{\vect}[1]{{\bm{#1}}}

\newcommand{\eq}[1]{\begin{equation}#1\end{equation}}
\newcommand{\eqs}[1]{\begin{equation}\begin{split}#1\end{split}\end{equation}}
\newcommand{\eqnref}[1]{Eq.\,\eqref{#1}}
\newcommand{\figref}[1]{Fig.\,\ref{#1}}
\newcommand{\tabref}[1]{Tab.\,\ref{#1}}

\usepackage{physics}
\newcommand{\rd}{\partial}
\newcommand{\change}[1]{{\color{black}#1}}

\begin{document}

\title{Signatures of a Deconfined Phase Transition on the Shastry-Sutherland Lattice: Applications to  Quantum Critical SrCu$_2$(BO$_3$)$_2$}

\author{Jong Yeon Lee}
\affiliation{Department of Physics, Harvard University, Cambridge, Massachusetts 02138, USA}

\author{Yi-Zhuang You}
\affiliation{Department of Physics, Harvard University, Cambridge, Massachusetts 02138, USA}
\affiliation{Department of Physics, University of California, San Diego, California 92093, USA}

\author{Subir Sachdev}
\affiliation{Department of Physics, Harvard University, Cambridge, Massachusetts 02138, USA}

\author{Ashvin Vishwanath}
\affiliation{Department of Physics, Harvard University, Cambridge, Massachusetts 02138, USA}

\date{\today}
\begin{abstract}
We study a possible deconfined quantum phase transition in a realistic model of a two-dimensional Shastry-Sutherland quantum magnet, using both numerical and field theoretic techniques. Using the infinite density matrix renormalization group (iDMRG)  method, we verify the existence of an intermediate plaquette valence bond solid (pVBS) order, with two fold degeneracy, between the dimer and N\'eel ordered phases. We argue that the quantum phase transition between the N\'eel and pVBS orders may be described by a deconfined quantum critical point (DQCP) with an {\em emergent O(4) symmetry}.
By analyzing the correlation length spectrum obtained from iDMRG, we provide evidence for the DQCP and emergent $\O(4)$ symmetry in the lattice model. Such a phase transition has been reported in the recent pressure tuned experiments in the Shastry-Sutherland lattice material  $\mathrm{SrCu_2 (BO_3)_2}$ \cite{Zayed2017}. The non-symmorphic lattice structure of the Shastry-Sutherland compound leads to extinction points in the scattering, where we predict sharp signatures of a DQCP in both the phonon and magnon spectra associated with the spinon continuum. The effect of weak interlayer couplings present in the three dimensional material is also discussed. Our results should help guide the  experimental study of DQCP in quantum magnets.
\end{abstract}
\maketitle

\tableofcontents

\section{Introduction}

Quantum magnets can host some of the most exotic phenomena in condensed matter physics, due to the strong quantum fluctuations of the microscopic spin degrees of freedom. Notably, deconfined gauge fluctuations and fractionalized spinon excitations can emerge in quantum magnets, which bear no analog in classical spin systems. Such behavior can exist either in a quantum spin liquid, which is a stable phase of matter with topological order \cite{FQH_Shot1_1997,FQH_Shot2_1997,FQH_Image_2004}; or by tuning a single parameter to a critical point known as the deconfined quantum critical point (DQCP) \cite{Senthil2004,Senthil2004_Science}. The DQCP describes the possible continuous phase transition between two distinct symmetry breaking phases, which is beyond the conventional Landau-Ginzburg paradigm. While the search for quantum spin liquids is still an ongoing research effort in condensed matter physics \cite{Review_Leon2017}, the possibility of observing the DQCP in materials could provide us with an alternative opportunity to study the properties of deconfined spinons and emergent gauge fields in quantum magnets\change{, as well as in interacting fermion systems that realizes the DQCP \cite{Fakher2016_DQCP, Fakher2017_DQCP_LL, You2018_bSPT_DQCP, Pollmann2018_DQCP, Ippoliti2018, Fakher2019_DQCP_qsh}.}

\begin{figure}[t]
\begin{center}
\includegraphics[width=0.85\columnwidth]{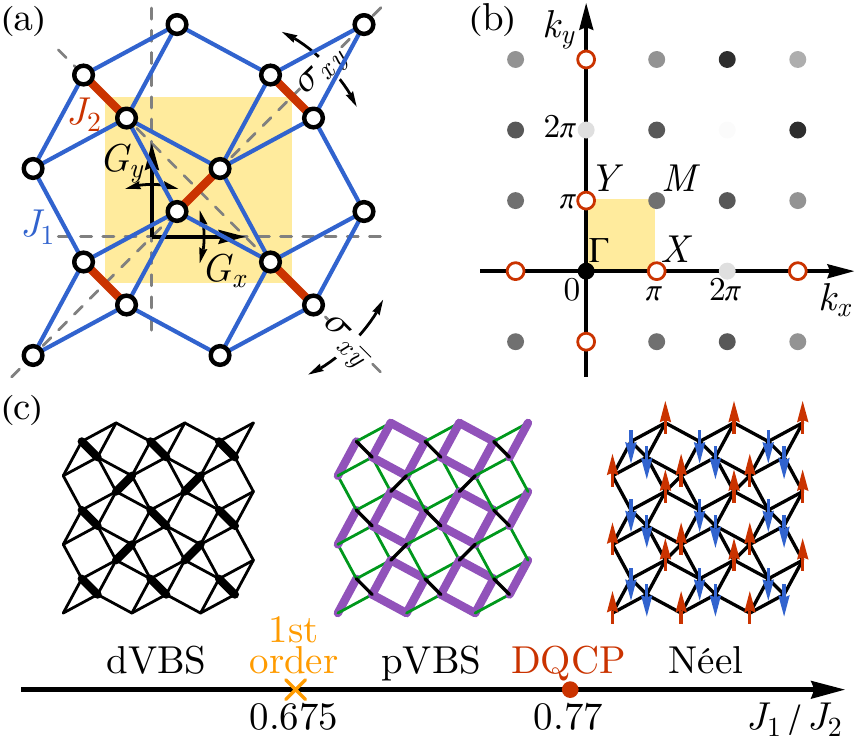}
\caption{(a) The Shastry-Sutherland lattice of copper sites (small circles) in $\mathrm{SrCu_2(BO_3)_2}$, on which the spins reside. The spins are coupled across nearest neighbor bonds ($J_1$, in blue) and dimer bonds ($J_2$, in red). Each unit cell contains four sites, as shaded in yellow. The glide reflection $G_x$, $G_y$ and the diagonal reflection $\sigma_{xy}$, $\sigma_{x\bar{y}}$ symmetries are indicated on the lattice. (b) Diffraction peaks from copper sites. The darker dot indicates a higher intensity. The extinction points are marked out by red circles. The first Brillouin zone is shaded in yellow, corresponding to the unit cell in (a). Special momentum points $\Gamma, X, Y, M$ are defined as labeled. (c) The phase diagram of the spin model \eqnref{eq:model}. The N\'eel antiferromagnetic and dimer valence bond solid (dVBS) phases are separated by the intermediate plaquette valence bond solid (pVBS) phase upon tuning the $J_1/J_2$ ratio. The critical points are determined in \tabref{tab:critical_regime} based on our iDMRG result. The transition between pVBS and N\'eel phases is likely to be a DQCP (or weakly first-order proximate to a DQCP).}
\label{fig:setup}
\end{center}
\end{figure}

In a recent experiment \cite{Zayed2017}, a phase transition between N\'eel antiferromagnet and plaquette valence bond solid \change{(crystal)} \cite{Mambrini2006} was observed in a single crystal of $\mathrm{SrCu_2 (BO_3)_2}$ under pressure. The material is a layered quantum magnet. Within each two-dimensional layer, the copper ions carry the spin-1/2 degrees of freedom and are arranged on a Shastry-Sutherland lattice as shown in \figref{fig:setup}(a). The spin system was proposed to be effectively described by the Shastry-Sutherland model \cite{ShastrySutherland1981,SSmodel1999}
\begin{equation}\label{eq:model}
H=J_1\sum_{ij\in\text{n.n.}}\vect{S}_i\cdot\vect{S}_j+J_2\sum_{ij\in\text{dimer}}\vect{S}_i\cdot\vect{S}_j,
\end{equation}
where the $J_1$ and $J_2$ bonds are specified according to \figref{fig:setup}(a). The ratio $J_1/J_2$ between the coupling constants is tunable by pressure in experiments within certain range. In the large $J_1$ (or large $J_2$) limit, the model reduces to the square lattice Heisenberg model (or the decoupled dimerized model), which stabilizes the N\'eel phase (or the dimer valence bond solid (dVBS) phase). Between these two limits, numerical \cite{Koga2000Qu,Lauchli2002Ph,Mambrini2006,Lou2012St,Corboz2013} and theoretical \cite{Chung2001} analysis of the model have revealed an intermediate plaquette valence bond solid (pVBS) phase, as illustrated in \figref{fig:setup}(c).  Remarkably, the experiment in Ref.\onlinecite{Zayed2017} seems to confirm this phase diagram. 
Since the pVBS and N\'eel phases separately break two distinct symmetries, the lattice and the spin rotation symmetry, a direct second-order transition between them would necessarily go beyond the Landau-Ginzburg paradigm and point to the possibility of the DQCP. Although the nature of the pVBS-N\'eel transition remains unresolved by experiments, there are promising signs for the exciting opportunity that $\mathrm{SrCu_2 (BO_3)_2}$ might provide the first experimental platform to realize DQCP.

Recent studies on different models with the same symmetry class showed that the transition between pVBS and N\'eel phases could be first-order\cite{Sandvik2018O4, Nahum2018O4}. However, despite being first-order, the transition is accompanied with an extended region of quantum-critical-like scaling and an emergent $\O(4)$ symmetry, implying that the transition could be close to a DQCP (possibly as an avoided criticality). Thus the DQCP is still the best theory to account for these anomalous features in the critical region, even though it may eventually break down at longest scales. Note that the $J$-$Q$ model or loop model studied in Monte Carlo simulations\cite{Sandvik2018O4, Nahum2018O4} are designed differently from the original Shastry-Sutherland model to avoid the sign problem. Given that the first- or second-order nature of the transition can be tuned by model parameters \cite{Sandvik_first_2010,YQQin2017,Zhang2018Co,Jian2017Em} and is therefore a model-dependent property, the fate of the pVBS-N\'eel transition in the Shastry-Sutherland model remains to be fully resolved yet.

The goal of this work is to investigate the pVBS-N\'eel transition in the Shastry-Sutherland model \eqnref{eq:model} in more detail using both field theory and the density matrix renormalization group (DMRG) approach, and to identify the unique signatures of DQCP that can be probed by inelastic neutron scattering (INS) or resonant inelastic X-ray scattering (RIXS) experiments. 
We use the infinite DMRG technique to overcome the sign problem. Our numerical simulation indicates (i) that the transition between pVBS and N\'eel phases appears continuous up to the largest available system size (infinite cylinder with the circumference of 10 lattice sites), although we can not rule out the possibility of a weakly first-order transition due to our limited system size. (ii) We also observe the asymptotic degeneracy between spin-triplet and spin-singlet excitations over a large length scale, demonstrating an approximate emergent $\O(4)$ symmetry which rotates among the N\'eel and pVBS order parameters. Our theoretical analysis further suggests that (iii) in the Shastry-Sutherland lattice, in contrast to previous realizations of DQCP,  a  dangerously irrelevant operator is absent which has consequences for numerics and that (iv) critical spinon continua  appear at the extinction points of lattice diffraction peaks (c.f.~\figref{fig:setup}(b)) in both the magnon and phonon channels at low-temperature around the DQCP. The universal critical behaviors of these continua are examined as well, which could guide the experimental study of the candidate DQCP in the $\mathrm{SrCu_2(BO_3)_2}$ material.


The rest of the paper is organized as follows. In Sec.~\ref{sec:DMRG}, we perform an infinite DMRG simulation on the Shastry-Sutherland spin model and discuss the nature of the phase transition between N\'eel and VBS phases based on a correlation length spectra. In Sec.~\ref{sec:sym_analysis}, we analyze symmetry quantum numbers of a monopole operator whose proliferation induces the transition to the VBS phase. By investigating the transformation property of the monopole, we show that the single-monopole term is suppressed while the double-monopole term can appear in the action describing the N\'eel order in the Shastry-Sutherland lattice. We compare the differences among various microscopic models -- easy-plane, rectangular, and Shastry-Sutherland -- whose possible emergent symmetry is all $\O(4)$. One distinct feature of the Shastry-Sutherland lattice is the presence of the relevant anisotropy operator that breaks the four-fold lattice rotation symmetry, which stabilizes the VBS order and gives rise to a fast-growing spectral gap in the spin-0 channel as the system enters the VBS phase. In Sec.~\ref{sec:spectral}, we propose the spectral signatures of DQCP in the Shastry-Sutherland model, including the $\SO(4)$ conserved current fluctuation in the magnon spectrum and the VBS fluctuation in the phonon spectrum. Both of these features appear at the extinction point of the Shastry-Sutherland lattice, which is detectable at low energy without overwhelmed by the elastic scattering signals. We conclude our discussion in Sec.~\ref{sec:discussion}.

\section{Numerical Study}\label{sec:DMRG}

In this section, we study the model in \eqnref{eq:model} using the infinite density matrix renormalization group (iDMRG) method \cite{White1992,McCulloch2008}. We wrap the 2D system onto a cylinder which is infinite along $x$-direction but compact along $y$-direction, with a finite circumference $L$. 
In the simulation, spin-1/2s are mapped into the hard-core bosons with density $\expval{n} = 1/2$ per site; an antiferromagnetic spin interaction would then be translated into hopping and density-density interaction terms for bosons. As the boson number is conserved in the simulation, we have an explicit $\U(1)_z$ symmetry which allows us to extract correlation functions for an operator with a specific $\U(1)_z$ quantum number. During the simulation, we fixed the value of $J_2 = 1$ and tuned the value of $J_1$ across the phase transition between the pVBS and N\'eel order phases. Since the unit cell of Shastry-Sutherland lattice contains $2 \times 2$ square unit cells, the iDMRG unit cell becomes $2 \times (2m)$ slice of the infinite cylinder, where the circumference of the cylinder $L = 2m$.

In the iDMRG simulation, we have two limiting factors to describe the exact two-dimensional state: the circumference length $L$ and the bond dimension $\chi$. Due to limited computational capacity, it is difficult to conclusively identify whether the phase transition is continuous or weakly first order in the DMRG simulation. Still, we can extract useful information of the ground state by simulating the model with increasing values of $L$ and $\chi$. For the discussionon bond dimension scaling, see Appendix.\,\ref{app:bond_dimension}. In the DMRG simulation, we measured (i) energy, (ii) plaquette order parameter, and (iii) correlation length spectra.

\subsection{Detection of the pVBS Phase}

\begin{figure} [t]
 \includegraphics[width=0.47\textwidth]{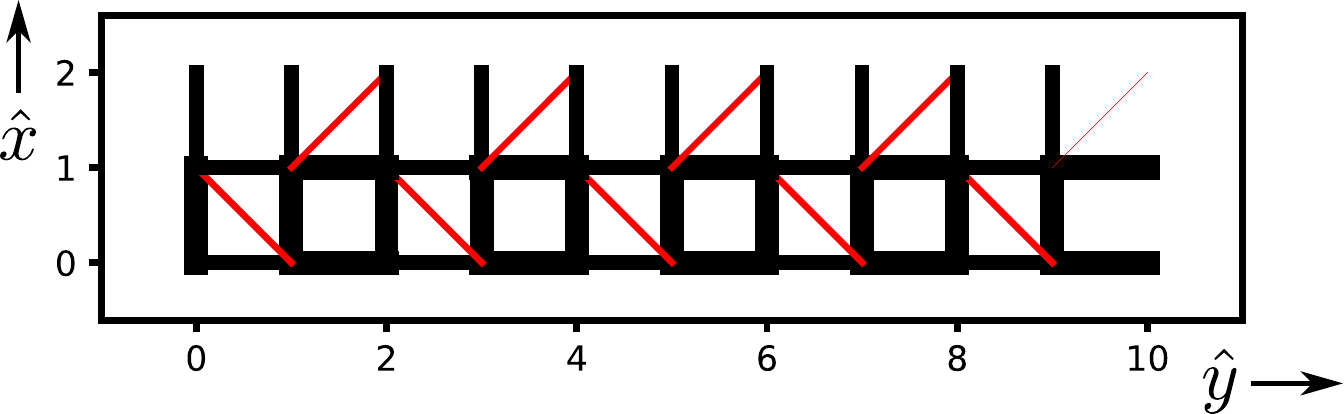}
 \vspace{-4pt}
\caption{\label{fig:pVBS_phase} Visualization of bond strengths $\expval{\vect{S}_i \cdot \vect{S}_j}$ within the iDMRG unit cell. $\hat{x}$-direction is along the cylinder, and $\hat{y}$-direction is around the circumference. The width of the bond represents the value of $\vect{S}_1 \cdot \vect{S}_2$. The line color is black (red) if $\vect{S}_1 \cdot \vect{S}_2$ is negative (positive). One can notice that the singlet is formed at a plaquette.  }
\end{figure}

Although the matrix product state description of the state is exact in the infinite bond dimension limit, for a finite bond dimension, the cylindrical geometry of the iDMRG simulation provides some bias to the preferred entanglement structure for the ground state. As a result, in the pVBS phase, one of two symmetry broken phases would be automatically chosen and the order parameter would not vanish in the iDMRG simulation. To characterize the pVBS phase, we define the plaquette order parameter as follows (See  \eqnref{eq:order_parameters} and \eqnref{eq:monopole_VBS} in Sec.~\ref{sec:sym_analysis} for a more rigors definition),
\begin{equation}
\Im\langle\scM\rangle\sim\sum_{i} (-)^x\langle \vect{S}_i\cdot\vect{S}_{i+\hat{x}}\rangle-(-)^y\langle \vect{S}_i\cdot\vect{S}_{i+\hat{y}}\rangle.
\end{equation}
In the pVBS phase, the dimer strengths $\langle\vect{S}_i\cdot\vect{S}_j\rangle$ on each bond $\langle ij\rangle$ can be visualized in \figref{fig:pVBS_phase}, which clearly demonstrates the pattern of the pVBS ordering around empty (square) plaquettes. Indeed, we measured that $\Im\langle\scM\rangle \neq 0$ and $\Re\langle\scM\rangle = 0$ in the paramagnetic regime as expected for the pVBS phase. (See \figref{fig:corr_spectra})

\begin{figure}
 \hspace{-7pt} \includegraphics[width=0.49\textwidth]{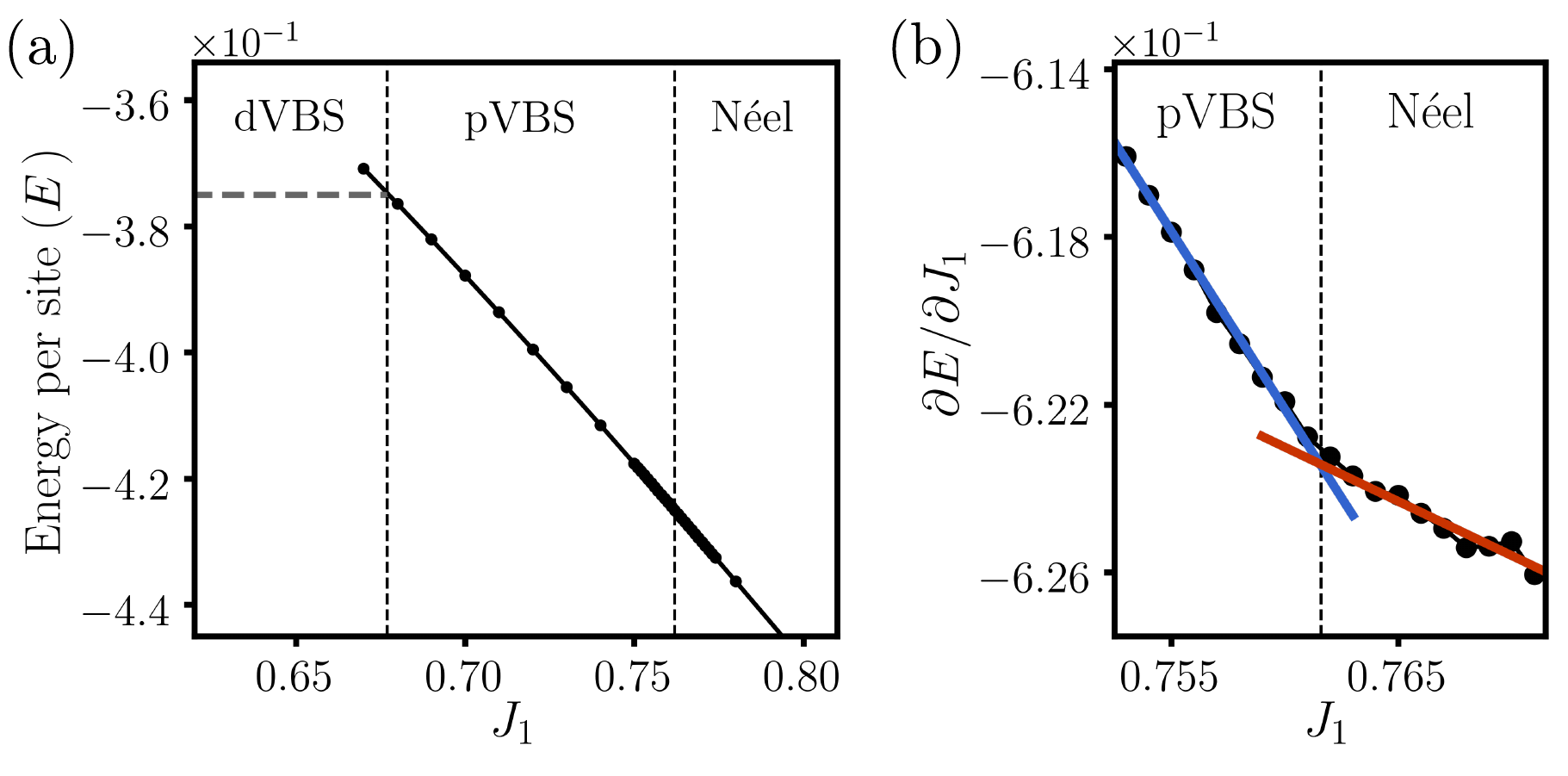}
 \vspace{-20pt}
 
\caption{\label{fig:energy_plot}At $L=10$, $\chi=4000$. (a) Energy ($E$) per site. The horizontal dotted line represents the exact ground state energy of dimer VBS state, $E_\text{site} = -0.375$. (b) $\rd E/\rd J_1$ per site near the transition between plaquette VBS and N\'eel order. The continuous first order derivative is a characteristic of the continuous phase transition. Black dashed lines denote phase boundaries among dimer VBS, plaquette VBS, and N\'eel order. }
\end{figure}

Although the plaquette order parameter $\Im\langle\scM\rangle$ is a useful indicator, it is not precise because (i) the system size does not reach the thermodynamic limit and (ii) the geometry of the iDMRG simulation provides some bias toward a specific entanglement structure for the ground state. Albeit small, it has a non-vanishing value in the N\'eel phase, see \figref{fig:corr_spectra}. Thus, we use a discontinuity in the second derivative of the energy $\rd^2 E/\rd J_1^2$ to locate the pVBS-N\'eel transition point. Up to $L=10$, the first order derivative of energy is continuous across the phase transition, implying that the transition is either a weakly first order or second order transition.
On the other hand, from the energy plot in \figref{fig:energy_plot}, the dVBS-pVBS transition point can be easily extracted because the dVBS state is an exact ground state of Eq.~\ref{eq:model} with energy per site $E_\textrm{site} = -0.375 J_2$. As we can see, the first order derivative of the energy is discontinuous here, signaling a clear first-order phase transition between the two spin singlet dVBS and pVBS phases. For $L<6$, we do not observe the pVBS phase. Transition points for different system sizes are summarized in \tabref{tab:critical_regime}.

\begin{table}[h]
\begin{center}
\begin{tabular}{c|c|c|c|c}
& $L = 6$ & $L= 8$ & $L = 10$ & $L=12$ \\
\hline\hline
\,$(J_1/J_2)_{c1}$ \,& \, 0.682 \,&\, 0.677 \,&\, 0.675 \,&\, 0.675 \,\\
$(J_1/J_2)_{c2}$ & 0.693 & 0.728 & 0.762 & 0.77
\end{tabular}
\end{center}
\caption{\label{tab:critical_regime} $(J_1/J_2)_{c1}$ ($(J_1/J_2)_{c2}$) is the transition point between the dVBS and pVBS (pVBS and N\'eel) phases. Transition points are extracted from the peak of the energy derivative at $\chi=4000$. At $L=12$, the DMRG simulation do not converge for different initial states at $\chi=4000$, resulting in different transition points. Thus, the critical point is determined as the midpoint between two transition points obtained from the VBS-like and N\'eel-like initial states.}
\end{table}


\subsection{Correlation Length Spectra, Monopole Fluctuations, and Emergent $O(4)$ Symmetry}

Here, we present a signature of the DQCP in the correlation length spectrum data obtained from the iDMRG simulation. 
A continuous phase transition is characterized by the divergence of a correlation length, and an equal-time correlation function exhibits the power-law decaying behavior $\langle\scO(r) \scO(0)\rangle \propto r^{-2 \Delta_{\cal O}}$ where $\Delta_{\cal O}$ is a scaling dimension of an operator ${\cal O}$. In particular, if there exists an emergent symmetry, operators unified under the emergent symmetry should share the same scaling dimension $\Delta_{\cal O}$ \cite{Nahum2015SO5}.  For example, at the conventional DQCP with the emergent $\SO(5)$ symmetry, the N\'eel $\vect{n}=(n_x,n_y,n_z)$ and VBS $\vect{v}=(v_x,v_y)$ order parameter should have the same power-law behavior:
\begin{equation}
    \expval{\vect{n}(r) \cdot \vect{n}(0)} \sim \expval{\vect{v}(r) \cdot \vect{v}(0)}  \sim 1/r^{1+\eta},
\end{equation}
where $\eta = 2 \Delta_{\cal O} - 1$ is the anomalous exponent defined relative to the engineering exponent in (2+1)D which is $1$. However, in the iDMRG simulation, the finite circumference $L$ of the cylinder and finite bond dimension $\chi$ introduce a cutoff length scale \cite{MPS_review}, which prevents us from observing the power-law behavior in the correlation function. Therefore, at long distances along the cylinder, the DMRG correlation function of an operator ${\cal O}$ will always decay exponentially with certain finite correlation length $\xi_{\cal O}$ in the simulation. Nevertheless, instead of trying to compare the scaling dimension $\Delta_\scO$, we can determine the emergent symmetry by comparing the correlation lengths  $\xi_\scO$ between N\'eel (spin-1) and VBS (spin-0) excitations. 
In Zauner et al. \cite{Zauner2015}, it has been found that the correlation length spectrum is inversely proportional to the energy of the excitations that mediate this correlation behavior. 
Thus, the correlation length spectra give access to the individual dynamics of different types of excitations.

\begin{figure}[t]
 \hspace{-7pt}\includegraphics[width=0.47\textwidth]{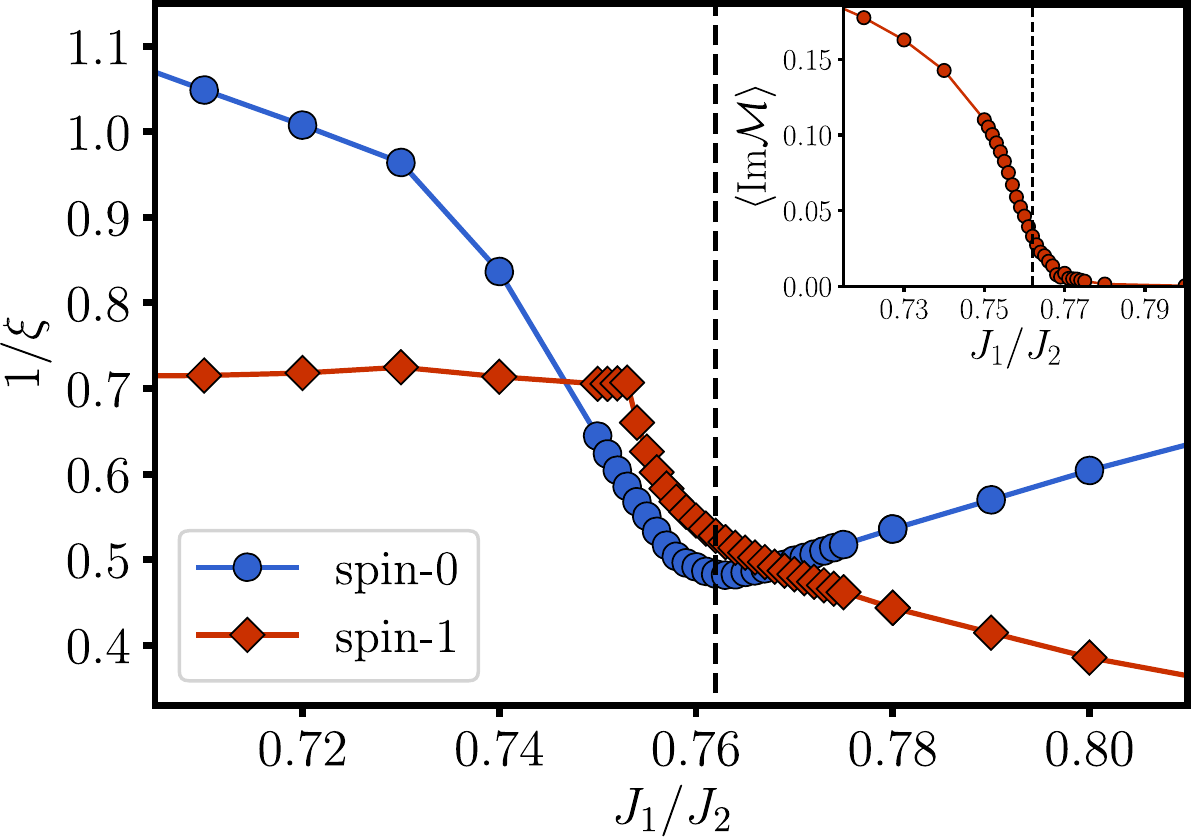}
\caption{\label{fig:corr_spectra} The inverse of the largest correlation length for spin singlet and triplet operators as a function of $J_1/J_2$ at $L=10$ and $\chi=4000$. Here, the dashed line represents the transition point extracted from the second order energy derivative. The small plot at the top right shows the plaquette order parameter ($\Im \cal M$) across the transition. Deep in the VBS phase, the correlation length of a spin-triplet operator is larger than that of a spin-singlet operator as expected by a mean-field theory. As we approach the critical point, we can observe that the correlation length of the spin-singlet sector becomes larger than that of the spin-triplet sector. This behavior agrees with what is expected from the scenario in Fig.~\ref{fig:DQCP_scenario}. 
}
\end{figure}

Using the DMRG transfer matrix technique, one can readily obtain the correlation lengths $\xi$ along the cylinder. Moreover, since our DMRG simulation has an explicit $\U(1)_z$ symmetry, extracted correlation lengths are labeled by $S_z$ quantum numbers. Since the microscopic model has full $\SO(3)$ symmetry, there will be an exact degeneracy among correlation lengths with different quantum numbers. For example, a three-fold degeneracy among $S_z = 0, \pm 1$ would imply that this correlation length corresponds to the excitation carrying the quantum number $S=1$ of the $\SO(3)$ symmetry. In this way, we can identify the $\SO(3)$ spin quantum number of each operator appeared in the correlation length spectrum.

In \figref{fig:corr_spectra}, we plot the inverse of the largest correlation lengths for spin-singlet and triplet operators. The spin-singlet operator corresponds to the pVBS order parameter (or the monopole operator $\scM$) at low-energy. The monopole operator is gapped in both the N\'eel and pVBS phases and only become gapless at the critical point. Therefore, the divergence of the spin-singlet correlation length can signal the onset of DQCP in the thermodynamic limit. 
Indeed, we identified that the peak of the singlet correlation length exactly coincides with the critical point extracted from the singularity of the energy derivative (Fig.~\ref{fig:energy_plot}). Furthermore, the correlation length $\xi_{S=0}$ at the critical point increases as the bond dimension $\chi$ increases. 
Although here we present only the correlation length spectrum at $L=10$ and $\chi=4000$, we also performed numerical simulations for different system sizes and obtained the result that the critical point summarized in \tabref{tab:critical_regime} coincides with the peak of the spin singlet correlation length. 
Therefore, we can infer that the transition is induced by the proliferation of monopoles, consistent with DQCP physics.

Moreover, to contrast our simulation result in the Shastry-Sutherland lattice with the conventional $\O(3)$ Wilson-Fisher transition between  an explicitly dimerized VBS and N\'eel order phases \cite{O(3)transition_1}, we performed the iDMRG simulation for a 2D $J_1$-$J'_1$ model with the antiferromagnetic Heisenberg coupling $J_1$ on nearest neighbor bonds together with the coupling  $J'_1$ on a fixed set of dimer covering bonds (See Appendix.~\ref{app:numerics}), and hence a unique VBS pattern is pinned by $J'_1$. Indeed, for this model, one can observe that the correlation length of the spin-triplet sector is always larger than the correlation length of the spin-singlet sector across the phase transition. This agrees with the picture discussed in Ref.~\cite{VBS_Sachdev,VBS_Campbell}, where the transition is triggered by the condensation of spin-triplet excitations (triplon) from the VBS phase. Within the mean-field theory framework, it was shown that the energy of the spin-triplet excitation $E_\textrm{triplon}$ is smaller than the energy of the spin-singlet excitation $E_\textrm{singlon}$ throughout the whole transition. Under the further assumption that the  singlon and triplon have the similar characteristic velocity $v$ \cite{Zauner2015}, one expect the aforementioned ordering of correlation lengths $\xi \sim (v/E)$. Thus, the inversion of the magnitude of correlation lengths for spin singlet and triplet operators at the transition signifies the unconventional feature of the DQCP in the Shastry-Sutherland lattice. For a detailed analysis regarding DQCP physics, see  Sec.~\ref{sec:dangerous}.

Finally, we remark that the spin-singlet correlation length $\xi_{S=0}$ and the  spin-tiplet correlation length $\xi_{S=1}$ approaches each other at the (finite size) critical point as we increase the system size. At $L=6$ and $\chi=4000$, the ratio $\xi_{S=1}/\xi_{S=0}\big|_\textrm{crit} = 0.33$, but at $L=10$ and $\chi=4000$, $\xi_{S=1}/\xi_{S=0}\big|_\textrm{crit} = 0.95$. From this trend, we expect to have $\xi_{S=1}/\xi_{S=0}=1$ in the thermodynamic limit, which indicates that the spin-singlet and spin-triplet excitations will become degenerate at the critical point, forming the four-component vector representation of a larger $\O(4)$ symmetry group. 
Put differently, we can observe that the crossing point of $\xi_{S=0}=\xi_{S=1}$ in \figref{fig:corr_spectra} approaches to the critical point as we increase the system size, consistent with the emergent $\O(4)$ symmetry relating N\'eel and VBS order parameters.


\section{Symmetry Analysis} \label{sec:sym_analysis}

The field theory of DQCP, the so called ``non-compact'' $\text{CP}^1$ theory, has the following form \cite{Motrunich:2004hh,Senthil2004}
\begin{equation} \label{eq:CP1}
    {\cal L}_\text{CP$^1$} = \abs{(\partial - ia ) {z} }^2 + \kappa \qty( \nabla \times a)^2 + \dots,
\end{equation}
where a two-component complex spinon ${z}=(z_1,z_2)^T$ is coupled to $\U(1)$ gauge field $a$. On top of this critical theory, one can have additional terms depending on symmetry of the system. The Shastry-Sutherland lattice has a $p4g$ space group symmetry, as shown in \figref{fig:setup}(a). The lattice respects two glide reflection $G_x, G_y$ and two diagonal reflection $\sigma_{xy},\sigma_{x\bar{y}}$ symmetries as illustrated in \figref{fig:setup}(a). The glide reflections and the (spinful) time-reversal $\scT$ symmetries can combine into composite symmetries $\scT G_x, \scT G_y$, dubbed as the \emph{time-reversal glide} symmetries. Note that glide-reflection symmetry is also broken in the N\'eel phase, while the time-reversal glide is not. Therefore, relative to the N\'eel phase, it is proper to think about the pVBS phases as to break the time-reversal glide symmetries.  

\begin{figure}[t]
\begin{center}
\includegraphics[width=0.25\textwidth]{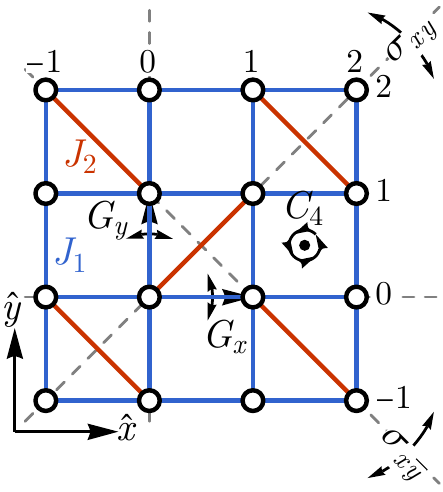}
\caption{The ideal Shastry-Sutherland lattice, deformed from \figref{fig:setup}(a). Each site $i$ can be labeled by a Cartesian coordinate $(x,y)$. The $x$ and $y$ coordinates are calibrated on the top and right respectively. The displacement vectors $\hat{x}=(1,0)$ and $\hat{y}=(0,1)$ are defined to connect nearest neighbor sites. $J_1$ and $J_2$ couplings are assigned to the nearest neighbor (in blue) and dimer (in red) bonds.}
\label{fig:square}
\end{center}
\end{figure}

To define the symmetry transformations more conveniently, we consider an ideal version of the Shastry-Sutherland lattice on a regular square lattice without distortion, as shown in \figref{fig:square}. This does not change the symmetry group but allows us to label every site by the Cartesian coordinate $(x,y)$ conveniently (where $x,y\in\dsZ$). The length of the nearest Cu-Cu bond is set to 1, such that the unit cell is of the size $2\times2$ (and hence the lattice constant is 2 here). With this, we can define the glide reflections $G_x:(x,y)\to(x+1,-y)$ and $G_y:(x,y)\to(-x,y+1)$, the diagonal reflections $\sigma_{xy}:(x,y)\to(y,x)$ and $\sigma_{x\bar{y}}:(x,y)\to(-y+1,-x+1)$, as well as the translations $T_x:(x,y)\to(x+2,y)$ and $T_y:(x,y)\to(x,y+2)$. Together, they generate the $p4g$ space group. The $p4g$ space group also contains a 90$^\circ$ rotation symmetry $C_4:(x,y)\to(-y+2,x-1)$ with respect to the center of the plaquette without a diagonal bond.  

On each site $i=(x,y)$, we define the spin operator $\vect{S}_i=(S_i^x,S_i^y,S_i^z)$, whose symmetry transformations are listed in \tabref{tab:symmetry} (assuming the spin rotation is not locked to the spatial rotation in lack of the spin-orbit coupling). The time-reversal symmetry $\scT$ is also included, which can flip all spin components. The N\'eel order parameter $\vect{n}=(n_x,n_y,n_z)$ and VBS order parameters $\vect{v}=(v_x,v_y)$ are defined as
\begin{equation}\label{eq:order_parameters}
\begin{split}
&\vect{n}\sim (-)^{x+y} \vect{S}_i, \\
&v_x\sim (-)^x(\tfrac{1}{4}-\vect{S}_i\cdot\vect{S}_{i+\hat{x}}), \\
&v_y\sim (-)^y(\tfrac{1}{4}-\vect{S}_i\cdot\vect{S}_{i+\hat{y}}),
\end{split}
\end{equation}
where each prefactor translates into the momentum $(\pi,\pi)$, $(\pi,0)$ and $(0,\pi)$ respectively in $k$-space, as marked out in \figref{fig:setup}(b), and the displacement vectors are defined to be $\hat{x}=(1,0)$ and $\hat{y}=(0,1)$. The VBS order parameters $v_x,v_y$ can be combined to form the following operator given a certain gauge choice (Appendix.\ref{app:monopole})
\begin{equation}\label{eq:monopole_VBS}
\scM^\dagger=\frac{1}{\sqrt{2}}\big((v_x+v_y)+\ii(v_x-v_y)\big),
\end{equation}
known as the monopole operator in the literature \cite{Senthil2004, Levin2004}, which corresponds to a hegehog-monopole of the N\'eel order parameter $\vect{n}$ in the spacetime. The monopole event changes the skyrmion number of the $\vect{n}$ field configuration by $+1$, and it is equivalent to inserting $2\pi$-flux of $\U(1)$ gauge field $a$ in the CP$^1$ theory \eqnref{eq:CP1}. The symmetry properties of $\vect{n}$ and $\scM^\dagger$ follows from those of the spin operator $\vect{S}_i$ and is summarized in \tabref{tab:symmetry}. For a detailed derivation, see Appendix~\ref{app:monopole}. Apart from these discrete symmetries, there is also an $\SO(3)$ spin rotation symmetry, under which $\vect{n}$ transforms as a $\SO(3)$ vector.

\begin{table}[t]
\begin{center}
\begin{tabular}{c|c|cc|c|c}
 & $(k_x,k_y)$ & $\vect{n}$ & $\scM^\dagger$ & $\vect{S}_i$ & $f_i$\\
\hline\hline
$G_x$ & $(k_x,-k_y)$ & $-\vect{n}$ & $-\scM^\dagger$ & $\vect{S}_{G_x(i)}$ & $(-)^yf_{G_x(i)}$\\
$G_y$ & $(-k_x,k_y)$ & $-\vect{n}$ & $-\scM^\dagger$ & $\vect{S}_{G_y(i)}$ & $f_{G_y(i)}$\\
$\sigma_{xy}$ & $(k_y,k_x)$ & $\vect{n}$ & $\scM$ & $\vect{S}_{\sigma_{xy}(i)}$ & $(-)^{xy}f_{\sigma_{xy}(i)}$\\
$\sigma_{x\bar{y}}$ & $(-k_y,-k_x)$ & $\vect{n}$ & $\scM$ & $\vect{S}_{\sigma{x\bar{y}}(i)}$ & $(-)^{x(y+1)}f_{\sigma_{x\bar{y}}(i)}$\\
\hline
$\scT$ & $(-k_x,-k_y)$ & $-\vect{n}$ & $\scM$ & $-\vect{S}_i$ & $\scK \ii\sigma^2 f_i$ \\
$\scT G_x$ & $(-k_x,k_y)$ & $\vect{n}$ & $-\scM$ & $-\vect{S}_{G_x(i)}$ & $(-)^y\scK \ii\sigma^2 f_{G_x(i)}$ \\
$\scT G_y$ & $(k_x,-k_y)$ & $\vect{n}$ & $-\scM$ & $-\vect{S}_{G_y(i)}$ & $\scK \ii\sigma^2 f_{G_y(i)}$
\end{tabular}
\end{center}
\caption{Symmetry transformation of momentum $(k_x,k_y)$,  N\'eel order $\vect{n}$, monopole operator $\scM^\dagger$, spin operator $\vect{S}_{i}$ and fermionic spinon $f_{i}$ (in the sense of PSG).}
\label{tab:symmetry}
\end{table}

\begin{figure}[t]
\begin{center}
\includegraphics[width=0.82\columnwidth]{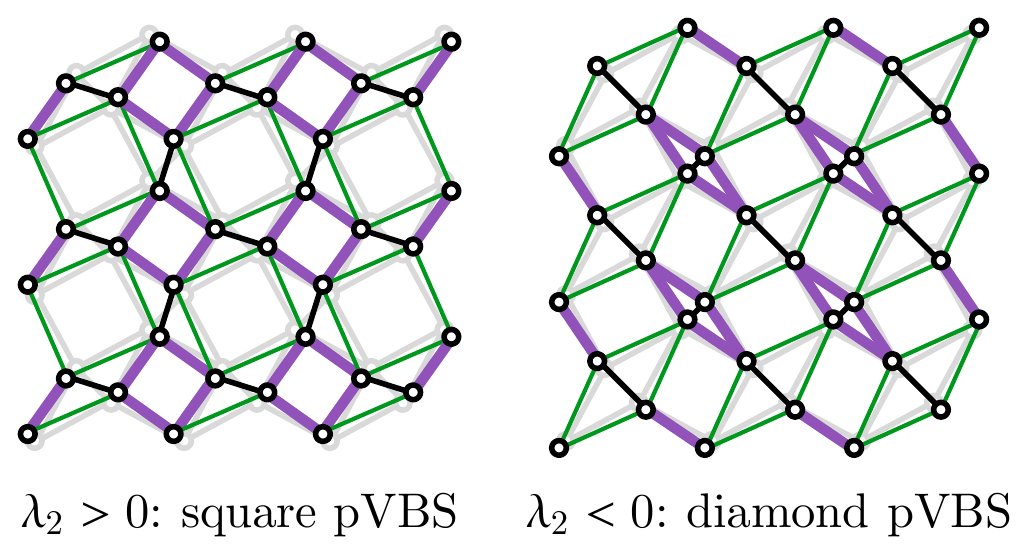}
\caption{Two types of plaquette valence bond solid (pVBS) phases. Thick purple links encircle the plaquette on which the spin singlet is formed. A diamond plaquette contains the diagonal bond, while a square plaquette does not. Each type of the pVBS order induces a corresponding lattice distortion. The undistorted lattice is shown as the background for contrast. Depending on the sign of $\lambda_2$, either diamond plaquette or square plaquette pVBS is favored.}
\label{fig:pVBS}
\end{center}
\end{figure}

The expectation value of the monopole operator $\langle\scM\rangle$ defined in \eqnref{eq:monopole_VBS} serves as a unified order parameter for various types of VBS orders. Depending on the phase angle of $\langle\scM\rangle$, the columnar VBS (cVBS) is described by $\langle\scM\rangle\sim\pm e^{\pm\ii \pi/4}$ and the plaquette VBS (pVBS) is described by $\langle\scM\rangle\sim\pm1$ (diamond plaquette)  or $\langle\scM\rangle\sim\pm\ii$ (square plaquette) as illustrated in \figref{fig:pVBS}. On the other hand, $\langle\scM\rangle\neq 0$ could also be interpreted as the condensation of monopoles. So the DQCP, as a transition from the N\'eel phase into the VBS phase, can be thought as driven by the VBS ordering or equivalently by the monopole condensation (starting from the N\'eel phase), which can be tuned by a monopole chemical potential $r$ in the Lagrangian as $r\scM^\dagger \scM$. The transition happens as $r$ changes sign. The condensation of monopole establishes the VBS order on the one hand and simultaneously destroys the N\'eel order on the other hand, due to a nontrivial topological term among the N\'eel and VBS order parameters, which was analyzed in details in Ref.\,\onlinecite{Senthil2004,Senthil2004_Science}. This scenario provides a plausible description of a direct continuous transition between the N\'eel and VBS phases.

However, apart from the apparent tuning parameter term $r\scM^\dagger \scM$, we must also include other symmetry-allowed (multi-)monopole terms in the Lagrangian, which could crucially influence the properties of the DQCP. On the Shastry-Sutherland lattice, to the leading order, they take the form of 
\begin{equation}\label{eq:LscM}
\scL_\scM=r \scM^\dagger \scM+\lambda_2\Re\scM^2+\cdots.
\end{equation}
Here we adopt the short-hand notations $\Re\scO=(\scO+\scO^\dagger)/2$ and $\Im\scO=(\scO-\scO^\dagger)/(2\ii)$ for generic operator $\scO$. Given the symmetry properties in \tabref{tab:symmetry}, one can see that the single monopole term, no matter $\Re \scM$ or $\Im\scM$, is forbidden by the glide reflection symmetry $G_x$ or $G_y$.
Furthermore, the imaginary part of the double-monopole term $\Im \scM^2$ is forbidden by the diagonal reflection symmetry $\sigma_{xy}$ or $\sigma_{x\bar{y}}$. Note that these symmetries exist in the critical theory as the spontaneous symmetry breaking has not yet occurred and the microscopic model has the symmetries. 


The higher-order monopole terms ($\scM^4, \scM^6, ...$) are expected to be less relevant and are therefore not included in \eqnref{eq:LscM} explicitly. Therefore the double-monopole term $\lambda_2\Re\scM^2$ is the most relevant monopole perturbation allowed on the Shastry-Sutherland lattice. Depending on its sign, the system will favor a square plaquette (or diamond plaquette) VBS order in the VBS phase, describe by the order parameter $\Im\scM$ (or $\Re\scM$), if $\lambda_2>0$ (or $\lambda_2<0$), as demonstrated in \figref{fig:pVBS}. The square and diamond pVBS orders have distinct symmetry properties. Under the reflection symmetries $\sigma_{xy}$ and $\sigma_{x\bar{y}}$, $\Im\scM \mapsto -\Im \scM$ while $\Re\scM$ stays invariant (see \tabref{tab:symmetry}), so the square pVBS spontaneously breaks the reflection symmetries while the diamond pVBS does not. Additionally, square-plaquette-centered $C_4$ rotation symmetry is spontaneously broken in the diamond pVBS, while it is not in the square pVBS.  Therefore the two different pVBS orders will lead to different lattice distortions that are symmetry-wise distinguishable in the experiments in the X-ray/neutron diffraction or NMR \cite{Takigawa2007, Takigawa2010No, Zayed2017}. 

Previous studies \cite{SS3D_Koga2000, Corboz2013} as well as our iDMRG data show that the pVBS phase has a square plaquette order. Thus $\lambda_2>0$ should be relevant to our discussion of the pVBS-N\'eel transition in the Shastry-Sutherland model \eqnref{eq:model}. $\lambda_2 > 0$ can be also argued based on the microscopic Hamiltonian in \eqnref{eq:model}. In the analysis of the pVBS phases, there are two types of singlet plaquette configurations: $s$-wave and $d$-wave types \cite{Mambrini2006} represented by the following singlet pairing configurations in a plaquette:
\begin{equation}
\quad\includegraphics[width=0.65\textwidth]{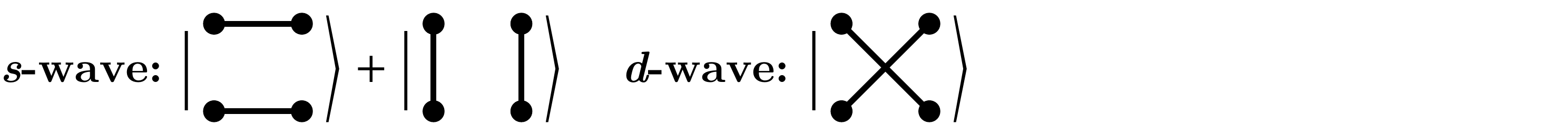} \vspace{4pt}
\end{equation}
For simplicity, consider a single plaquette with four spin-1/2s on the corners. 
For both square and diamond plaquettes, there is a AFM coupling $J_1$ along plaquette sides; for the diamond plaquette, there is an additional AFM $J_2$ coupling across one diagonal. Then, the $s$-wave and  $d$-wave singlet configuration has the energy $-2J_1 +1/4 J_2$ and $-3/4 J_2$ respectively for the diamond plaquette. As the pVBS phase exists at a parameter regime $J_1/J_2 \sim 0.7$, an estimation of the singlet configuration energy as listed in \tabref{tab:pVBS_energy} indicates that the $s$-wave pairing in the square plaquette has the lowest energy. In the iDMRG simulation, the wavefunction indeed exhibits the $s$-wave pairing symmetry in the pVBS phase. \change{A recent exact-diagonalization study \cite{Nakano2018Th} on the small cluster of spins ($N_s = 40$) also reported that the phase next to the N\'eel order phase hosts a spin-spin correlation which contradicts to the $d$-wave singlet in a diamond plaquette.}  Therefore it is natural to have $\lambda_2 > 0$ in the phenomenological field theory.

\begin{table}[bhtp]
\caption{Singlet configuration energy around different plaquette with different pairing symmetry, estimated from $J_1/J_2\sim0.7$ for the pVBS phase.}
\begin{center}
\begin{tabular}{c|c|c}
& $\qquad$ $s$-wave $\qquad$ &$\qquad$ $d$-wave $\qquad$\\
\hline \hline
square & $-2J_1\sim-1.4J_2$ &  0 \\
diamond & $-2 J_1 + 0.25 J_2\sim-1.15J_2$ & $-0.75J_2$ 
\end{tabular}
\end{center}
\label{tab:pVBS_energy}
\end{table}

\begin{figure}
\includegraphics[width=0.24\textwidth]{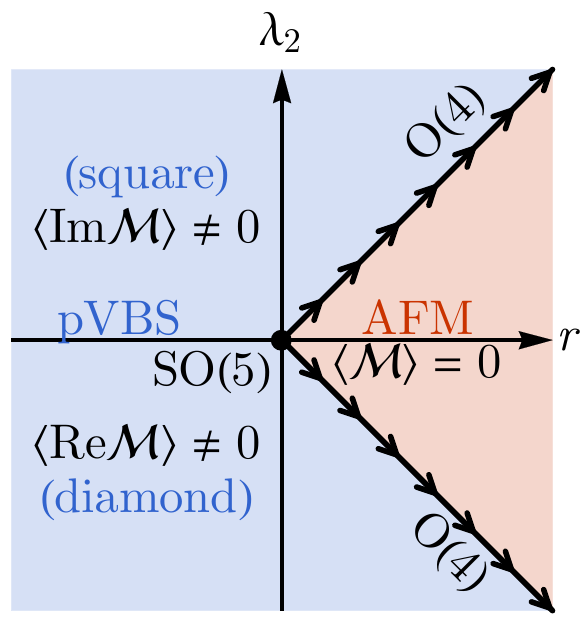}
\caption{\label{fig:DQCP_metlitski} Phase diagram of the appearance of $\O(4)$ DQCP on perturbing the $\SO(5)$ theory. Arrows indicate the RG flow direction.}
\end{figure}

At the first glance, it seems that the double-monopole term $\lambda_2$ in \eqnref{eq:LscM} is relevant and may destroy the DQCP. However, it is realized in Ref.\,\onlinecite{maxryan17} that $r$ and $\lambda_2$ actually recombine into a new tuning parameter $\tilde{r}$ and a new relevant perturbation $\tilde{\lambda}_2$. In the case of $\lambda_2>0$, the Lagrangian $\scL_\scM$ in \eqnref{eq:LscM} can be written as
\begin{equation}\label{eq:monopole_shastry}
\scL_\scM=\tilde{r}(\Im\scM)^2+\tilde{\lambda}_2(\Re\scM)^2+\cdots,
\end{equation}
with $\tilde{r}=r - \lambda_2$ and $\tilde{\lambda}_2=r+\lambda_2$. The parameter $\tilde{r}$ still drives a transition at $\tilde{r}=0$ (or equivalently $r=\lambda_2$), as shown in Fig.\ref{fig:DQCP_metlitski} with a modified emergent symmetry. At the transition point, the relevant perturbation $\tilde{\lambda}_2= 2 \lambda_2 > 0$ simply gaps out the diamond plaquette pVBS fluctuation $\Re\scM$ from the low-energy sector, leaving the square plaquette pVBS fluctuation $\Im\scM$ quantum critical. It is further argued that the pVBS fluctuation $\Im\scM$ will become degenerate with the N\'eel fluctuation $\vect{n}$ at the critical point\cite{maxryan17}, because the perturbations $\vect{n}^4$, $\vect{n}^2(\Im\scM)^2, (\Im\scM)^4, \cdots$ that can break the symmetry that rotates N\'eel and pVBS are all rank-four operators, which are expected to be irrelevant at the critical point. Therefore the N\'eel and VBS order parameters can combine into a $\O(4)$ vector $(\vect{n},\Im\scM)$, manifesting an emergent $\O(4)$ symmetry. The remaining topological $\O(4)$ $\Theta$-term still ensures that the development of the pVBS order $\Im\scM$ will simultaneously destroy the N\'eel order $\vect{n}$, establishing a direct pVBS-N\'eel transition with emergent $\O(4)$ symmetry.

\section{Dangerously Irrelevant Scaling and Its Absence}\label{sec:dangerous}

In this section, we discuss a peculiarity of the DQCP in the Shastry-Sutherland lattice compared to the other DQCP scenarios that have been extensively discussed.\cite{Senthil2004_Science,Motrunich:2004hh,Senthil2004,You2016St,maxryan17,Wang2017,Jian2017Em,Jian:2017qe} In the presence of the $\tilde{\lambda}_2$ term in \eqnref{eq:monopole_shastry}, the $\U(1)$ symmetry of the monopole operator (which acts as $\scM\to e^{\ii\theta}\scM$) is explicitly broken down to $\dsZ_2$ (at the lattice level). This $\dsZ_2$ symmetry can be identified as the glide reflection symmetry ($G_x$ or $G_y$) on the Shastry-Sutherland lattice, which can be further broken spontaneously in the pVBS phase by its order parameter $\expval{\Im\scM }$. At the DQCP, this $\dsZ_2$ symmetry will be restored and combined with the $\SO(3)$ spin rotation symmetry to form the larger emergent $\O(4)$ symmetry, denoted as $\SO(3)\times\dsZ_2\to\O(4)$. Although the $\dsZ_2$ symmetry is restored at the DQCP, it is never further enlarged to the $\U(1)$ symmetry of monopole conservation, because the explicit symmetry breaking term $\tilde{\lambda}_2$ is relevant. The presence of the relevant coupling $\tilde{\lambda}_2$ leads to an important difference between the DQCP on the Shastry-Sutherland lattice with the more conventional DQCP on the square lattice. 

To expose the differences and connections, let us briefly mention the other two lattices: the square lattice and the rectangular lattice. Due to the different lattice symmetries, the allowed leading monopole terms will be different, as summarized in \tabref{tab:lattice}. They will lead to different VBS orders and different properties of the DQCP. For example, on a rectangular lattice, the other double-monopole term $\lambda'_2\Im\scM^2$ is allowed but $\lambda_2\Re\scM^2$ is forbidden, which favors the horizontal/vertical cVBS order depending on $\lambda'_2>0$ (or $\lambda'_2<0$). The DQCP on the rectangular lattice has a similar emergent $\O(4)$ symmetry, which was carefully analyzed in Ref.\,\onlinecite{maxryan17}. However, on a square lattice, the four-fold rotational symmetry forbids all the double-monopole terms, leaving the quadruple-monopole term $\lambda_4\Re\scM^4$ most relevant, which favors cVBS (or pVBS) if $\lambda_4>0$ (or $\lambda_4<0$). In the absence of the double-monopole term, the DQCP on the square lattice has an even larger emergent symmetry $\SO(3)\times\dsZ_4 \to \SO(5)$. In the easy-plane model, the lattice symmetry would be that of the square lattice, but the spin-rotation symmetry is reduced from $\SO(3)$ to $\U(1)$. Here, the symmetry enhancement would be $\U(1)\times\dsZ_4  \to \O(4)$  \cite{YQQin2017,Wang2017}. 

\begin{table}[t]
\begin{center}
\begin{tabular}{c|c|c}
Global Symmetry & $\scL_\scM$ & Emergent Symmetry \\
\hline\hline
Square ($p4m$) & \, $\lambda_4 \Re \scM^4$ \, & $\SO(3)\times\dsZ_4\to\SO(5)$\\
Easy-plane Square & \, $\lambda_4 \Re \scM^4$ \, & $\U(1)\times\dsZ_4\to\O(4)$\\
Rectangular ($pmm$) & \, $\lambda'_2 \Im \scM^2$ \, & $\SO(3)\times\dsZ_2\to\O(4)$\\
Shastry-Surtherland ($p4g$) & \, $\lambda_2 \Re \scM^2$ \, & $\SO(3)\times\dsZ_2\to\O(4)$ 
\end{tabular}
\end{center}
\caption{\label{tab:lattice} Symmetry-allowed most-relevant monopole terms (apart from $r\scM^\dagger\scM$) and the corresponding DQCP emergent symmetries on different lattices}
\end{table}

\begin{figure}[t]
\includegraphics[width=0.48\textwidth]{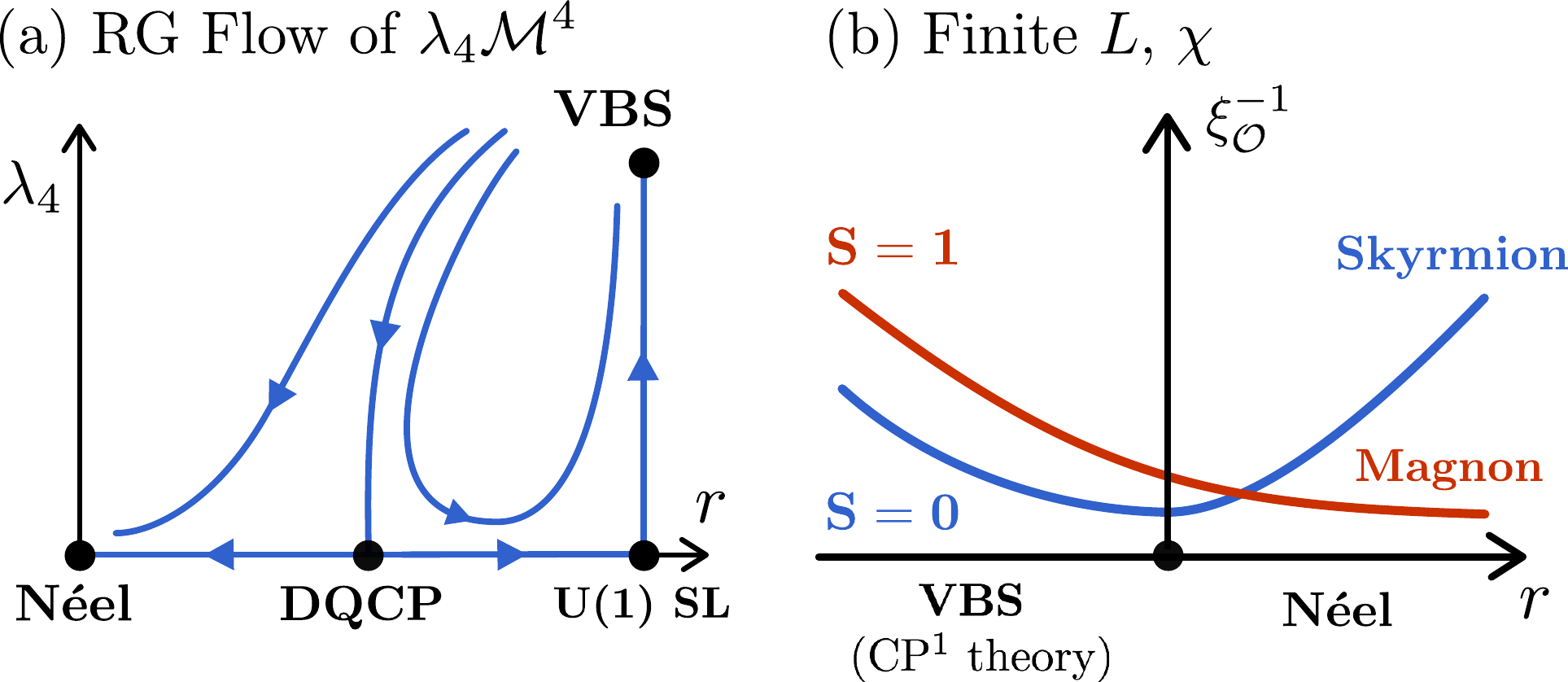}
\caption{ (a) Renormalization group (RG) flow of the quadrupled monopole term $\lambda_4 {\cal M}^4$ in the square lattice DQCP scenario, which is dangerously irrelevant. For a small deviation of the tuning parameter $J_1/J_2$ towards the VBS phase, $\lambda_4 $ initially decreases under RG flow until the RG scale reaches the spin-spin correlation length. Only after that, $\lambda_4 $ begins to increase to reach the VBS fixed point. This has noticeable consequences for observables in a finite size numerical simulation.
(b) Schematic plot for the inverse correlation length $\xi^{-1}$ of spin singlet and triplet operators as a function of tuning parameter for the square lattice DQCP scenario with either O(4) or SO(5) emergent symmetries. Note that $\xi^{-1}$ is related to the energy (mass gap) of the associated excitation \cite{Zauner2015}. Here, the skyrmion corresponds to the Higgsed `photon' excitation in the CP$^1$ theory. In the VBS phase, this photon excitation manifests as a spin singlet VBS order parameter fluctuation, i.e. the VBS domain wall thickness. Similarly, the magnon becomes a `triplon' in the VBS phase. The plot assumes the simulation of the DQCP at the finite circumference $L$ and bond dimension $\chi$, which prevents $\xi$ to diverge at the DQCP. 
\label{fig:DQCP_scenario}
}
\end{figure}


\begin{figure}[t]
 \includegraphics[width=0.43\textwidth]{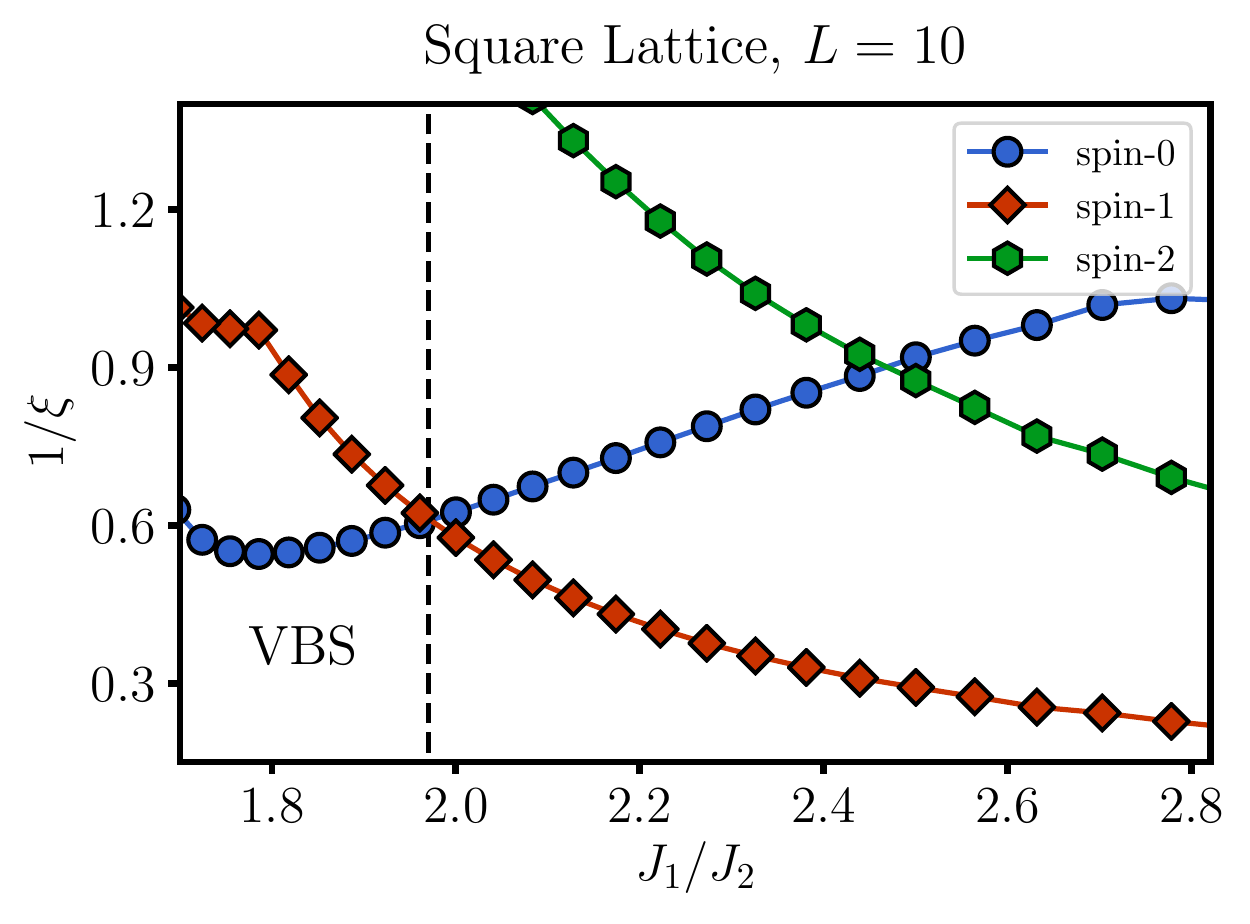} 
 
 \includegraphics[width=0.43\textwidth]{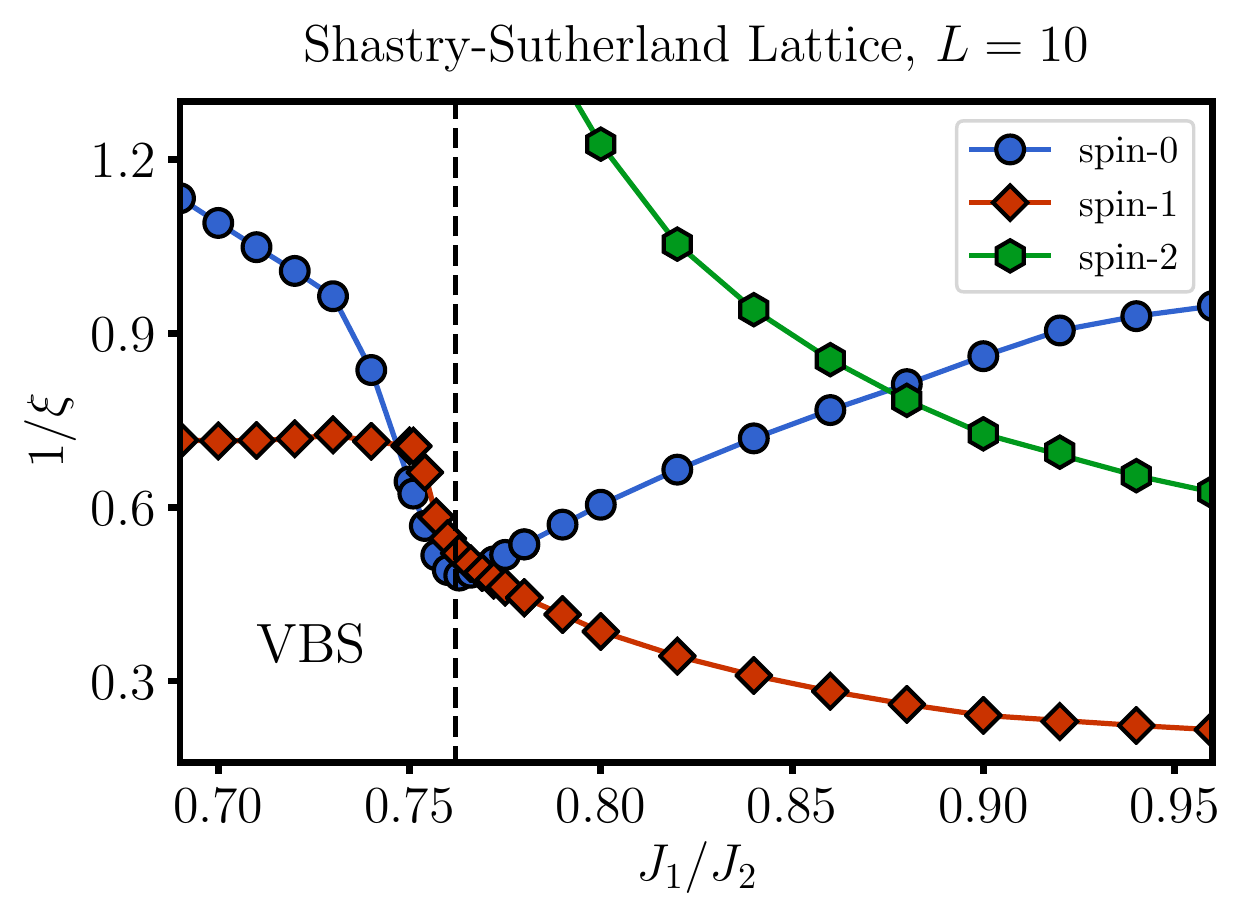}
\caption{ \label{fig:J1J2_spectra} The correlation length spectrum for (Left) $J_1$-$J_2$ model in the spin-1/2 square lattice ($\chi=2000$) and (Right) the Shastry-Sutherland model ($\chi=4000$). The correlation length spectrum coincides with the excitation level crossing spectrum in Fig.2 of \cite{WangSandvik2018} (After switching $J_2/J_1$ to $J_1/J_2$). For the value of $J_1/J_2$ smaller than the range shown in the figure, we would get the collinear striped AFM order (dVBS) for the square (Shastry-Sutherland) lattice. The level-crossing behaviors for both models are similar in the N\'eel ordered side.}
\end{figure}

Although the DQCP on the square lattice with the easy-plane deformation has the same $\O(4)$ emergent symmetry as the DQCP on the rectangular or Shastry-Sutherland lattice, there is a crucial difference between them.
On the square lattice, the $\U(1)\to\dsZ_4$ symmetry breaking term $\lambda_4\Re\scM^4$ is \emph{dangerously irrelevant}, which enhances $\dsZ_4$ to $\U(1)$ at the DQCP.
As we move away from the DQCP toward the VBS phase, the system will exhibit two different length scales: the spin correlation length $\xi_\text{spin}$ and the VBS domain wall width $\xi_\text{VBS}$ \cite{Senthil2004,Shao2016Qu}. The system is critical at the length scale below $\xi_\text{spin}$, meaning that the spin correlation decays in a power-law. Beyond this length scale, however, the system is still not fully in the VBS phase. Because the $\Re\scM^4$ operator is irrelevant at the critical point, its coupling coefficient $\lambda_4$ has decreased under the renormalization group (RG) flow initially, as in \figref{fig:DQCP_scenario}(a). However, once RG flow goes beyond the length scale of $\xi_\text{spin}$,  $\lambda_4$ starts to increase with the RG flow until it becomes strong enough to break the $\U(1)$ symmetry down to $\dsZ_4$ at the length scale $\xi_\text{VBS}$ \cite{Balents2007}. A careful analysis \cite{Senthil2004} showed that $\xi_\text{VBS} \sim \xi_\text{spin}^{(\Delta-1)/2}$, where $\Delta > 3$ (irrelevant) is the scaling dimension of $\Re\scM^4$ \footnote{Beyond the length scale $\xi_\text{spin}$, the system behaves like XY model with order parameter $\scM$. From the inspection of the XY model, the domain wall length scale is given by $\xi_\text{VBS} \sim \lambda_4^{-1/2}$, where $\lambda_4$ is a renormalized coefficient for $\Re\scM^4$ at that length scale. From the generic scaling argument, we have $\xi_\text{VBS} \sim L f(\lambda_4 L^{3-\Delta})$, where $L$ can be any length scale within the scaling regime. At the length scale $L=\xi_\text{spin}$ where the system behaves like XY model, we already know that $\xi_\text{VBS} \sim \lambda_4^{-1/2}$. Thus, we can deduce $f(x) \sim x^{-1/2}$, and derive that $\xi_\text{VBS} \sim \xi_\text{spin}^{(\Delta - 1)/2}$}.
Thus, $\xi_\text{VBS}$ grows more rapidly than $\xi_\text{spin}$ as we approach the critical point.

However, on the Shastry-Sutherland lattice, there is no symmetry enhancement from $\dsZ_2$ to $\U(1)$ at the DQCP because  $\U(1)\to\dsZ_2$ symmetry breaking term $\tilde{\lambda}_2(\Re\scM)^2$ is relevant. As a result, there is no dangerously irrelevant RG flow around the DQCP. As soon as we move away from the critical point, $\mathbb{Z}_2$ anisotropy is there to break the emergent $\O(4)$ symmetry down to the microscopic $\SO(3)\times\dsZ_2$ symmetry. Thus, there does not exist a separation of length scales nor scenario expected in \figref{fig:DQCP_scenario}. The singlet gap (the gap of pVBS fluctuation) should open up immediately as we tune $\tilde{r}$ away from the critical point in the thermodynamic limit.

What is the consequence of all these observations? 
In the study of finite size systems, the RG flow should stop at a certain point beyond the system size. This means that the dangerously irrelevant scaling can significantly affect the correlation behavior or excitation spectrum in small system sizes, as illustrated in \figref{fig:DQCP_scenario}(b). Due to the dangerously irrelevant scaling, near the DQCP towards the VBS side, the correlation length of the VBS order parameter fluctuation ($S=0$) should be  larger than the correlation length of the N\'eel order parameter fluctuation ($S=1$). 
However, in the thermodynamic limit, such a behavior is not guaranteed as the eventual fate under the RG flow is often difficult to understand. 
For example, in the mean-field treatment of the VBS phase, it has been argued that the spin-triplet excitation (triplon) has a lower energy than the singlet excitation (singlon) \cite{VBS_Campbell}. 
On the other hand, in the finite size systems, the dangerously irrelevant scaling \emph{enforces} the region with $\xi_{S=0}> \xi_{S=1}$ in \figref{fig:DQCP_scenario}(b) to appear regardless of the eventual RG behavior in the VBS phase. 
\change{Moreover, since we consider a quasi two-dimensional system in the iDMRG simulation, the finite circumference size can also affect the behavior in \figref{fig:DQCP_scenario}. In principle, the Mermin-Wagner theorem prevents the spontaneous symmetry breaking of a continuous symmetry in dimensions $D\geq 2$ \cite{MerminWagner}. In other words, in a quasi two-dimensional system, the strong fluctuation of the continuous order parameter (e.g. SO(3) spin-rotation) would gap out the system and reduce the correlation length for the associated Goldstone bosons (e.g. magnons). As the disordering effect would be stronger for the physical SO(3) spin-rotation symmetry than for the emergent $\U(1)$ symmetry enhanced from $\mathbb{Z}_4$, we would again expect the parameter region of $\xi_{S=0} > \xi_{S=1}$ to appear near the DQCP.}

By contrast, in the Shastry-Sutherland model, the relevant $\mathbb{Z}_2$ perturbation can always gap out the monopole fluctuation away from the critical point. To confirm this, we performed the iDMRG simulation on the spin-1/2 $J_1$-$J_2$ model at the same circumference size $L=10$ and compared the correlation length spectra between two models. In \figref{fig:J1J2_spectra}(a), we observe that $\xi_{S=0}$, the correlation length of a $S=0$ local excitation, is always larger than $\xi_{S=1}$ in the entire VBS phase on the square lattice. However, In \figref{fig:J1J2_spectra}(b), $\xi_{S=0}$ is larger than $\xi_{S=1}$ only at the transition point and immediately becomes smaller than $\xi_{S=1}$ as we tune the system away from the critical point towards the VBS phase. This is one of the non-trivial prediction from the presence of relevant anisotropy operator at a finite-system size simulation. In the N\'eel ordered phase, these two models exhibit very similar correlation length spectra. For more details, see Appendix.~\ref{app:numerics}.

Our numerical result for the square $J_1$-$J_2$ model aligns with the recent work \cite{WangSandvik2018} on the finite-DMRG simulation which calculated first several excited states with different spin quantum numbers. If we replace the excitation energy in Ref.~\cite{WangSandvik2018} with $\xi^{-1}$, we obtain the same crossing behavior. This can be justified by the fact that when a local excitation has a mass gap $m$, the correlation function mediated by the local excitation decays as $\sim e^{-m r}$, thus $m \propto \xi^{-1}$. Therefore, our theoretical scenario explains why a local \footnote{In the VBS phase, there are four degenerate ground states, whose degeneracy is lifted by the finite system size. These nearly degenerate excited states should be distinguished from the other excitations that can truly be `local'. } spin-singlet excitation in the Ref.~\cite{WangSandvik2018} has lower energy than a spin-triplet excitation around the critical point. It also elucidates the reason why the previous DMRG results of the $J_1$-$J_2$ model on the square lattice \cite{J1J2_Fisher2014, square_SL2013, Mambrini2006, square_pVBS2000, square_pVBS2003, square_IPEPS2009} were unable to identify the nature of the VBS phase without applying a pinning field. Because all previous numerics were also performed in the similar system size, they were in the regime where the VBS order parameter fluctuations were severe, disallowing one to confirm whether it is a plaquette or columnar VBS. Therefore, we conclude that the absence or presence of an irrelevant operator is essential to understand physical observables in any finite size system.

In summary, we note that previously the two promising scenarios to realize DQCP with spin 1/2 was either the SO(3) symmetric or easy plane deformations, both with four-fold degenerate VBS orders. Here, however, we have the two-fold degenerate VBS order. Nevertheless, as discussed in Ref.~\cite{maxryan17},  the two-fold monopole with SO(3) symmetry is equivalent to the easy plane deformation, if an enlarged SO(5) symmetry is assumed in the absence of these perturbations. 
Furthermore, we remark that the scaling behavior of dangerously irrelevant operator enables us to understand multiple observations made in previous numerical simulations which is absent in the Shastry-Sutherland model here.

\section{Spectral Signatures of DQCP}\label{sec:spectral}

A hallmark of the DQCP is the emergence of deconfined spinons at the critical point, which entails distinct features in both the magnon and  phonon excitation spectra that can be probed in INS or RIXS experiments. To better appreciate the predicted spectral features at low-temperature around the pVBS-N\'eel transition, we need to first understand the background elastic scattering signal of $\mathrm{SrCu_2(BO_3)_2}$ in its high-temperature paramagnetic phase without any symmetry breaking. We will focus on the scattering of neutrons or photons off of the copper sites. As shown in \figref{fig:setup}(a), there are four copper sites in each unit cell, coordinated at $\vect{r}_A=(1+\delta,1+\delta)/2$, $\vect{r}_B=(1-\delta,3+\delta)/2$, $\vect{r}_C=(3+\delta,1-\delta)/2$, $\vect{r}_D=(3-\delta,3-\delta)/2$ respectively, with the distortion parameter given by $\delta=0.544$ according to Ref.\,\onlinecite{SSmodel1999}. In the high-temperature paramagnetic phase, an elastic scattering experiment will reveal lattice diffraction peaks at a set of momenta $\vect{Q}=\pi(H,K)$ ($H,K\in\dsZ$, note that the lattice constant is 2 in our convention) with the amplitude given by $S(\vect{Q})=\int\dd^2\vect{r}\rho(\vect{r})e^{\ii\vect{Q}\cdot\vect{r}}\simeq\sum_{a=A,B,C,D}e^{\ii \vect{Q}\cdot\vect{r}_a}$, where $\rho(\vect{r})$ can represent either the electron density from Cu orbitals (which scatters X-ray photons) or the  density of Cu nuclear (which scatters neutrons). The corresponding intensity $|S(\vect{Q})|^2$ is plotted in \figref{fig:setup}(b). Notably, there are extinction points in the diffraction pattern, protected by the glide reflection symmetry. Note that the system at the DQCP has the glide-reflection symmetries $G_x$ and $G_y$ although both the N\'eel and pVBS phases do not. \footnote{The N\'eel ordered phase has the time-reversal glide ${\cal T} G_{x,y}$, while the pVBS phase has the time-reversal ${\cal T}$. At the DQCP, we have both symmetries, thus the composition ${\cal T} \circ {\cal T}G_{x,y} = G_{x,y}$ should be there.} 
The glide reflections $G_x$ and $G_y$ act as lattice translations by a half lattice constant followed by the reflections about the translation directions, as illustrated in \figref{fig:setup}(a). The fact that the density distribution $\rho(\vect{r})$ at equilibrium respects all lattice symmetries (including $G_x$ and $G_y$) implies that $\rho(x,y)=\rho(x+1,-y)=\rho(-x,y+1)$. They impose the following constraints on the scattering amplitude 
\begin{equation}\label{eq:glide_sym}
S(Q_x,Q_y)=e^{\ii Q_x}S(Q_x,-Q_y)=e^{\ii Q_y}S(-Q_x,Q_y),
\end{equation}
which implies the extinction of diffraction peaks at $\vect{Q}\in \pi(2\dsZ+1,0)$ or $\pi(0,2\dsZ+1)$, as marked out in \figref{fig:setup}(b). 
In general, $\rho(\vect{r})$ can describe the spatial pattern of any scatterers that interact with the probing particle (e.g. X-ray or neutrons).  \eqnref{eq:glide_sym} holds under the assumption that the scatterer field $\rho(\vect{r})$ is even (symmetric) under glide reflection symmetry. This constraint can be generalized to other type of scatterers including magnetic fluctuations at finite frequency. For example, the destruction of the scattering amplitude at these extinction points could extend to finite-frequency inelastic scattering, as long as no scatterer at that energy scale breaks the glide reflection symmetry. 
However, as we lower the temperature and approach the pVBS-N\'eel transition, certain glide-reflection-breaking excitations (meaning that the scatterer is odd under the glide reflection)  may emerge at low energy as part of the quantum critical fluctuation. Indeed, we will show that the emergent $\SO(4)$ conserved current fluctuation and the pVBS order fluctuation are examples of glide-reflection-breaking scatterers, which will become critical at the DQCP and ``light up'' the extinction points. They will provide unique  signatures of the DQCP that are also relatively easy to resolve in experiments, as there are no background scattering signals at the extinction points.

\begin{figure}[htbp]
\begin{center}
\includegraphics[width=0.6\columnwidth]{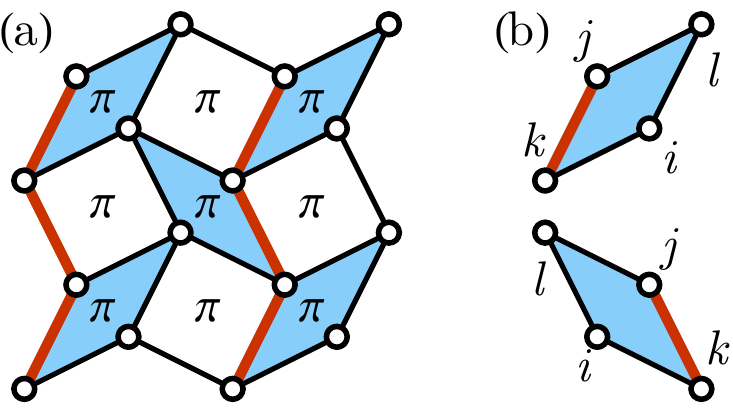}
\caption{(a) The $\pi$-flux model of fermionic spinons. The spinon hopping on the thick red bond gets a minus sign, such that each plaquette has a $\pi$-flux. Additional spinon interaction terms are applied to blue shaded diamonds. (b) The arrangement of the site indices $[ijkl]$ around each shaded diamond.}
\label{fig:spinon}
\end{center}
\end{figure}

The proposed spectral signatures at the pVBS-N\'eel transition is most convenient to analyze using a fermionic spinon theory for the DQCP, which has been shown to be equivalent (dual) to the conventional CP$^1$ theory\cite{youSMG2,Assaad2016,Ma2018Dy,Xu2018Mo}. In the fermionic spinon theory, the spin operator $\vect{S}_{i}=\frac{1}{2}f_{i}^\dagger \vect{\sigma} f_i$ is fractionalized into fermionic spinons $f_{i}=(f_{i\uparrow},f_{i\downarrow})^\intercal$, which are then placed in the $\pi$-flux state\cite{Rantner2002Sp,hermele2005,senthilfisher,ranwen} described by the following mean-field Hamiltonian
\begin{equation}\label{eq:HMF}
\begin{split}
H_{\text{MF}} =& -\sum_{\langle i j\rangle} t_{ij}(f_i^\dagger f_j+ \text{h.c.})\\
&+g\sum_{[ijkl]}(f_i^\dagger f_j+\text{h.c.})(f_k^\dagger f_k-f_l^\dagger f_l)+\cdots,
\end{split}
\end{equation}
where $\langle i j\rangle$ runs over all the nearest neighbor bonds and $[ijkl]$ runs over all the shaded diamond plaquette depicted in \figref{fig:spinon}(a). The spinon hopping amplitude $t_{ij}$ takes $t_{ij}=-t$ on the bonds highlighted in red in \figref{fig:spinon}(a) and $t_{ij}=t$ on the rest of the bonds, such that the spinon sees $\pi$-flux threading through each plaquette. The phenomenological parameter $t\sim J_1$ sets the energy scale of the spinon, which is expected to be of the same order as the spin interaction strength $J_1$. The $\pi$-flux state model \eqnref{eq:HMF} was originally proposed as an example of algebraic spin liquids\cite{Rantner2002Sp,hermele2005,senthilfisher,ranwen} (including the DQCP as a special case). It is recently confirmed via quantum Monte Carlo (QMC) simulations\cite{Assaad2016,Ma2018Dy,Xu2018Mo} that the $\pi$-flux state actually provides a pretty good description of the spin excitation spectrum at the DQCP.

The $\pi$-flux state mean-field ansatz determines a projective symmetry group (PSG)\cite{Wen:2002qr} that describes how the spinon should transform under the space group symmetry, as concluded in the last column of \tabref{tab:symmetry}. It is found that the spinon hopping along the dimer bonds are forbidden by the $\sigma_{xy},\sigma_{x\bar{y}}$ symmetry, which is not broken at the DQCP. Therefore the effect of $J_2$ can only enter the Hamiltonian as a four-fermion interaction to the leading order, given by the $g$ term in \eqnref{eq:HMF}. The $g$ term describes the spinon interaction around each diamond plaquette $[ijkl]$ where the site indices $i,j,k,l$ are arranged according to \figref{fig:spinon}(b). It turns out that the interaction $g$ is directly related to the $\lambda_2 \Re\scM^2$ term given that the monopole operator $\scM^\dagger\sim (v_x+v_y)+\ii (v_x-x_y)$ can be written in terms of VBS order parameters $v_x,v_y$, which are further related to fermionic spinons via $v_x\sim(-)^x f_{i+\hat{x}}^\dagger f_i+\text{h.c.}$ and $v_y\sim(-)^{x+y} f_{i+\hat{y}}^\dagger f_i+\text{h.c.}$, such that the interaction can be written as $g v_x v_y\sim g \Re\scM^2$. Therefore $g>0$ would correspond to $\lambda_2>0$, which favors the square plaquette VBS order $\Im\scM$. 

Let us ignore the interaction $g$ for a moment. By diagonalizing the spinon hopping Hamiltonian, we find the spinon dispersion $\epsilon_{\vect{k}}=\pm 2t\sqrt{\cos^2k_x+\cos^2k_y}$, which gives rise to 4 Dirac fermions (or equivalently 8 Majorana fermions) at momentum $(\pi/2,\pi/2)$. Note that the distortion parameter $\delta$ does not affect the spinon band structure. Naively, the spinon mean-field theory has an emergent $\SO(8)$ symmetry rotating among the 8 low-energy Majorana fermions. However, a $\SU(2)$ gauge structure must be introduced with the fractionalization, which reduces the emergent symmetry to $\SO(5)$. The interaction term $g$ plays an important role to further break the $\SO(5)$ symmetry explicitly to $\O(4)$, matching the emergent symmetry observed in the DMRG simulation.

\begin{figure}[t]
\begin{center}
\includegraphics[width=\columnwidth]{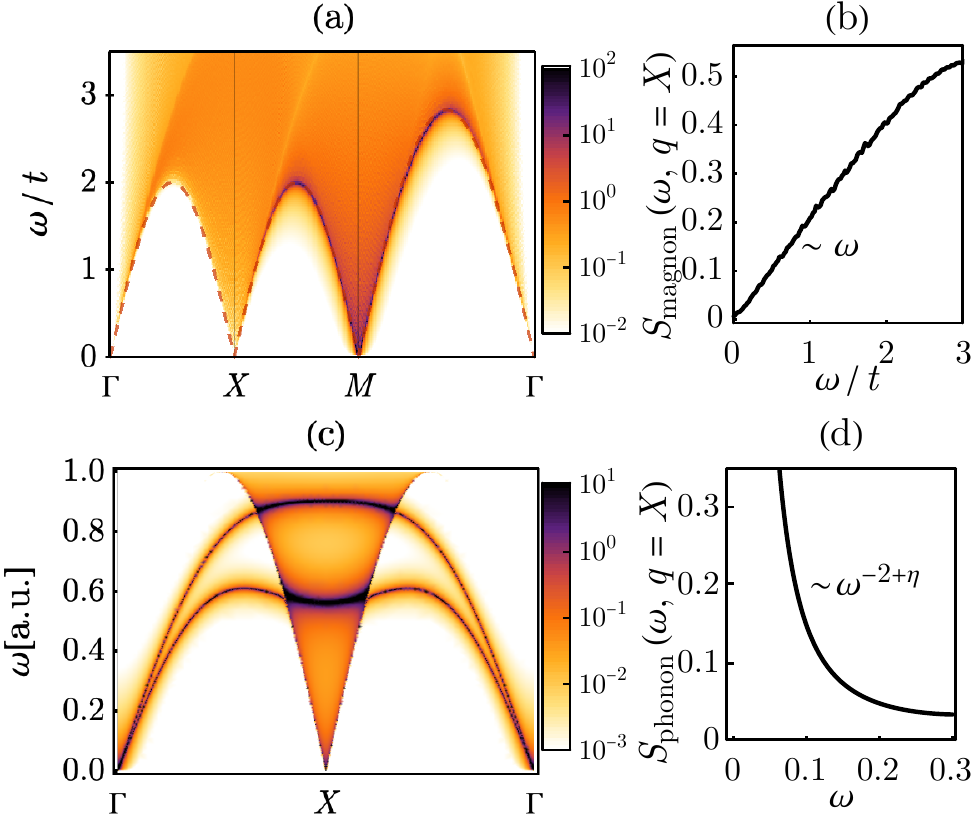}
\caption{(a) Spin excitation spectrum (dynamical structural factor) $S_\text{magnon}(\omega,\vect{q})$ at the DQCP. Darker color indicates higher intensity. The dashed line trace out the lower-edge of the continuum which is described by \eqnref{eq:omega_min}. (b) The frequency dependence of the spectral intensity at the extinction point $X$. At the low-frequency limit, the spectral intensity grows with frequency linearly, which manifests the conserved current associated to the emergent $\O(4)$ symmetry. (c) Schematic illustration of the phonon spectrum $S_\text{phonon}(\omega,\vect{q})$. The bare phonon dispersion is inferred from Ref.\,\cite{Haravifard2012NSFSDLSSSGS}. The VBS-phonon coupling leads to a continuum in the phonon spectrum. (d) The frequency dependence of the phonon spectral intensity at the extinction point $X$. The intensity falls off in a power-law with frequency, whose exponent should be the same as that of the spin fluctuation at $M$ point.}
\label{fig:spectrum}
\end{center}
\end{figure}

Based on the fermionic spinon mean-field ground state, the spin excitation spectrum $S(\omega,\vect{q})$ has been calculated in Ref.\,\onlinecite{Assaad2016,Ma2018Dy,Xu2018Mo} and is reproduced in \figref{fig:spectrum}(a) for illustration, where the high symmetry points $\Gamma,X,M$ are defined in \figref{fig:setup}(b). The lower edge of the spinon continuum is given by $\omega_\text{min}(\vect{q})=\min_{\vect{k}}|\epsilon_{\vect{k}+\vect{q}}-\epsilon_{\vect{k}}|$, which reads
\begin{equation}\label{eq:omega_min}
\omega_\text{min}(\vect{q})=2t\sqrt{\sin^2q_x+\sin^2q_y},
\end{equation}
as shown in \figref{fig:spectrum}(a). This provides us a way to estimate the mean-field parameter $t$ from the experimentally measured spin excitation spectrum, by fitting this lower edge. This is a rather robust result as its shape is unaffected by the distortion parameter $\delta$. 

Remarkably, a gapless continuum will appear on top of the extinction point $X$ (as well as $Y$ by $\sigma_{xy}$ symmetry) as seen in \figref{fig:spectrum}(a). As previously analyzed in \eqnref{eq:glide_sym}, the spin excitation at the $X$ point must be odd under the glide reflection symmetry $G_x$. This symmetry constraint enforces that the gapless continuum should correspond to the emergent $\SO(4)$ conserved current fluctuation $\vect{J}_{y}=\vect{n}\partial_y \Im\scM-\Im\scM\partial_y\vect{n}$ which involves both the N\'eel $\vect{n}$ and pVBS $\Im\scM$ order parameters and has been thoroughly studied in Ref.\,\onlinecite{hermele2005,Ma2018Dy,Ma2018No}. Since $\vect{n}$, $\partial_y$ and $\Im\scM$ are all odd under the glide reflection $G_x$ (that preserves the extinction point $X$), the current operator $\vect{J}_y$ is also odd under $G_x$. Therefore \eqnref{eq:glide_sym} does not hold anymore, such that the fluctuation of $\vect{J}_y$ can appear at the extinction point $X$ as a spinon continuum in the magnon channel (as $\vect{J}_y$ carries spin-1). On the other hand, fluctuations like $\vect{n}\Im\scM$ is not allowed to appear at the $X$ point in the spin excitation spectrum because the bound state $\vect{n}\Im\scM$ is even under glide reflection (cf. \eqnref{eq:glide_sym}).  The $\vect{J}_y$ continuum will only become gapless if both the N\'eel and pVBS fluctuations are gapless, which only happens at the DQCP. Protected by the emergent $\O(4)$, the conserved current operator must have a scaling dimension exactly pinned at 2, which indicates that the magnon spectral weight at the extinction point $X$ must increase with the frequency linearly (at below the energy scale of $J_1$)
\begin{equation}
S_\text{magnon}(\omega,\vect{q}=X)\propto \omega.
\end{equation}
as shown in \figref{fig:spectrum}(b). We propose this as a hallmark feature of the spin fluctuation at the pVBS-N\'eel transition in $\mathrm{SrCu_2(BO_3)_2}$. Confirmation of this linear frequency growth of the spectral weight will provide direct evidence for the emergent $\O(4)$ symmetry at the DQCP.

Apart from the features in the spin excitation (magnon) spectrum, the DQCP also introduces new features to the phonon spectrum, due to the pVBS-phonon coupling. The pVBS order has a linear coupling to the lattice displacement as its representation under the lattice symmetry group matches with that of strain fields
\begin{equation}
\scL_{\text{VBS-phonon}}\sim v_x u_x + v_y u_y,
\end{equation}
where $u_x$ (or $u_y$) is the lattice displacement in the $x$ (or $y$) direction with a momentum $(\pi,0)$ (or $(0,\pi)$). It is crucial that $u_x$ and $u_y$ here are not acoustic phonon modes around momentum $(0,0)$ \change{in a continuum theory}, otherwise they can only enter the field theory in the form of derivatives $\partial_i u_j$. The coupling is allowed by lattice symmetry and is evident from the pVBS induced lattice distortion as demonstrated in \figref{fig:pVBS}(a). This leads to a hybridization between phonon and pVBS fluctuations. As the pVBS fluctuation becomes critical (gapless) at the DQCP, the quantum critical fluctuation will also appear in the phonon spectrum due to the hybridization effect, as illustrated in \figref{fig:spectrum}(c) (see Appendix \ref{app:phonon} for detailed analysis). As the pVBS order carries the momentum $(\pi,0)$ and $(0,\pi)$, the phonon continuum will also get softened at these momenta, which happen to be the extinction points $X$ and $Y$ of the lattice diffraction pattern. Note that the pVBS fluctuation is odd under glide reflection, therefore new phonon continuum is allowed to emerge at the extinction points with the spectral weight diverging at low-frequency following a power-law,
\begin{equation}\label{eq:Sphonon}
S_\text{phonon}(\omega,\vect{q}=X)\propto\omega^{-2+\eta.}
\end{equation}
The anomalous dimension $\eta$ should match that of the pVBS order parameter at the DQCP, which, by the emergent $\O(4)$ symmetry, is also the same $\eta$ of the N\'eel order parameter. Based on the QMC simulations in Ref.\,\onlinecite{YQQin2017,XFZhang2017}, $\eta$ has been estimated to be $\eta=0.13\sim0.3$. Observation of such critical phonon fluctuations at the extinction points with a power-law divergent spectral weight as shown in \figref{fig:spectrum}(d) will be another direct evidence of DQCP.

In conclusion, extinction points are protected by the glide reflection symmetry, but both the conserved current and the pVBS order parameter breaks the glide reflection. Their critical fluctuations are therefore allowed to appear at the extinction point. This is rather a nice property that there will be no background signals form lattice diffraction, which makes these spectral signatures of DQCP more easier to resolve in experiments. Our analysis indicates that the conserved current fluctuation (which carries spin-1) should appear in the magnon spectrum and the pVBS fluctuation (which carries spin-0) should appear in the phonon spectrum. Observation of these critical fluctuations in the scattering experiment will be strong evidence for the potential realization of DQCP in $\mathrm{SrCu_2(BO_3)_2}$.

\section{Effects of Interlayer Coupling}

\begin{figure}[t]
 \hspace{-5pt} \includegraphics[width=0.4\textwidth]{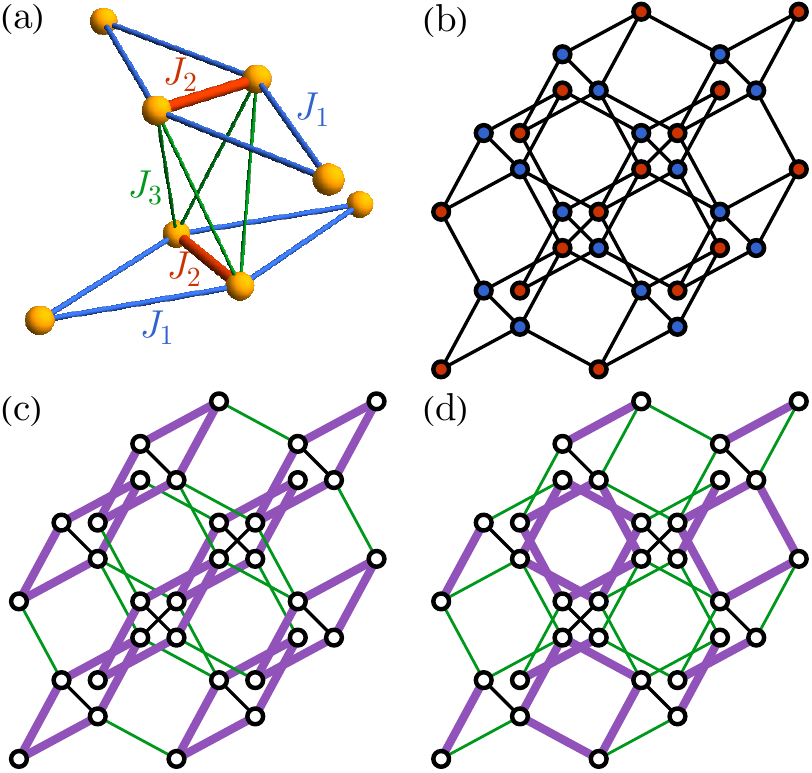}
\caption{ \label{fig:3D_coupling} (a) Three-dimensional interlayer AFM coupling $J_3$  among spins located at the cross between upper and lower diagonal bonds, mediated by the interpenetrating oxygen atom. The other interlayer couplings are negligible. 
Preferred stacking of (b) the N\'eel ordered phase with $\vect{n}^{(l)} = \vect{n}^{(l+1)}$ (red and blue dots for opposite spins), (c) and the diamond pVBS phase with $\Re \scM^{(l)} = \Re \scM^{(l+1)}$, (d) square pVBS phase with $\Im \scM^{(l)} = -\Im \scM^{(l+1)}$ (thick purple bonds mark singlet plaquettes). For (c) and (d) cases, energetically favorable stacking can be deduced by the interlayer dimer resonance.}
\end{figure}

So far, we have discussed the DQCP physics assuming that the system is two-dimensional. However, the material realization of \eqnref{eq:model}, SrCu$_2$(BO$_3$)$_2$, has a three-dimensional structure which is a stack of the Shastry-Sutherland lattices with a relative shift given by the lattice vector (1,1) between neighboring layers. Each layer is separated by the layer of oxygen, but due to the super-exchange term mediated by the oxygen, there is a small interlayer anti-ferromagnetic interaction $J_3 \sim 0.1 J_2$ \cite{Ueda2003} between the spin-1/2s located at the crossed dimers in \figref{fig:3D_coupling}. { Since the stacking structure preserves $G_{x,y}$ and $\sigma_{xy,x\bar{y}}$ symmetries, previous monopole analysis still holds for each two-dimensional layer. } Here, we would like to better understand the effect of the three-dimensional interlayer coupling to the DQCP physics.

To analyze, we first consider coupling layers of spin systems at the conventional DQCP point with the emergent $\SO(5)$ symmetry (other than the $\O(4)$ case in the previous discussion). Here, each layer is described by the following nonlinear sigma (NLSM) model in Euclidean spacetime \cite{senthilfisher, Tanaka2005}:
\begin{equation}
    {\cal S}^{(l)}_\textrm{DQCP} = \int d^3 x \frac{1}{2g} (\partial_\mu \Phi ^{(l)} )^2 + 2\pi\ii\Gamma_\textrm{WZW}[\Phi^{(l)}]
\end{equation}
where $\Phi^{(l)} = (\vect{n}_x, \vect{n}_y, \vect{n}_z, \Im \scM, \Re \scM)$ is the order parameter of the $l$-th layer and $\Gamma_\textrm{WZW}[\Phi^{(l)}]$ is a Wess-Zumino-Witten (WZW) term at level $k=1$.  With the interlayer coupling, the total action is given by the following general form \cite{Bi:2015qv}
\begin{equation}
    {\cal S} = \sum_l {\cal S}^{(l)}_\textrm{DQCP} - \sum_l\int d^3 x \, g^{ab} \Phi_a^{(l)} \Phi_b^{(l+1)},
\end{equation}
where the interlayer coupling coefficients $g^{ab}$ can be arranged into a matrix $g$. Given that the scaling dimension of $\Phi$ is estimated to be around $\Delta_\Phi\simeq 0.6$ in the literature\cite{Lou2009,Harada2013,Block2013,Pujar2013,Sreejith2014,Sreejith2015,Pujari2015,Dyer2015,Nahum2015a,Nakayama2016}, the interlayer coupling is expected to be relevant (as $2\Delta_\Phi<3$), locking the order parameters across different layers. 
Then, the coefficient matrix $g$ becomes important since the sign of its determinant $\det g$ crucially affects how the topological (WZW) term from each layer will be added up (cf.\,\cite{Bi:2015qv}).  More precisely, if the sign of $\det g$ is positive (negative), the WZW terms are added up in a uniform (staggered) manner. The reason is that we can always redefine $\Phi^{(l)}$ in alternate layers to make $g$ positive definite, but the price to pay will be that the WZW term will change sign alternatively if the original $\det g$ is negative. \change{Note that the procedure requires the topological term to be invariant under the change of origin (translation symmetric), i.e. $T_{x,y}: \Gamma_\textrm{WZW}[\Phi] \mapsto \Gamma_\textrm{WZW}[\Phi]$. Otherwise, the addition or subtraction of topological terms across  layers would not be well-defined due to the arbitrariness of the choice of origin across the layers for the locked order parameter.}

For example, consider coupling the two-dimensional square lattice $J$-$Q$ models \cite{Sandvik2007} into a 3D cubic lattice with AFM spin-spin interaction along vertical bonds. Each layer putatively realizes the DQCP physics with $\Phi = (\vect{n},v_x, v_y)$ describing N\'eel and cVBS order parameters. Under the AFM interlayer coupling, the coupling matrix $g$ has the sign structure of $g^{aa} = (-,-,-,+,+)$. This is because the vertical AFM coupling prefers $\vect{n}^{(l)} = - \vect{n}^{(l+1)}$ while the vertical plaquette ring exchange \footnote{The AFM spin-spin interaction along vertical bonds induces the effective vertical plaquette ring exchange for the low energy subspace.} favors $v_{x,y}^{(l)}=v_{x,y}^{(l+1)}$.  In this case, $\det g < 0$, so the WZW terms in neighboring layers tend to cancel each other. However the cancellation will not be exact, as the $\Phi$ field still admits (smooth) fluctuation over layers, this results in a residual topological term, namely a topological $\Theta$-term. Staggering a WZW-term at level $k$ would give a $\Theta$-term at $\Theta = \pi k$. Now the problem of the coupled DQCPs boils down to understand the fate of the $\SO(5)$ NLSM with $\Theta=\pi$ in (3+1)D. There are some hints from the fermionic parton analysis. One can consider fractionalizing the $\Phi$ vector to the bilinear from of a fermionic parton field $\psi$ following $\Phi_a\sim\bar{\psi}\ii\gamma^5\Gamma^a\psi$, such that the $\psi$ fermion is in the $\SO(5)$ spinor representation (or the $\Sp(2)$ fundamental representation). The emergent gauge structure will be $\SU(2)$, which points to the $\SU(2)$ quantum chromodynamics (QCD) model in (3+1)D with $\Sp(2)$ flavor symmetry,
\begin{equation}
    \scL=\bar{\psi}(\gamma^\mu D_\mu+m
    +\ii\gamma^5\Phi_a\Gamma^a)\psi.
\end{equation}
Integrating out the fermion and gauge fluctuation is expected to reproduce the $\SO(5)$ NLSM with $\Theta=\pi(1+\sgn m)$, such that $\Theta=\pi$ is realized at $m=0$. However, the number of Dirac fermion flavors ($N_f=2$) is not enough to avoid a chiral symmetry breaking in 3D. Therefore, under the interlayer coupling, it is likely that the $\SO(5)$ DQCP flows to a discontinuous transition point induced by the quantum fluctuation. \change{Considering that the $\O(4)$ DQCP from easy-plane anisotropy or rectangular deformation  is descended from the $\SO(5)$ DQCP \cite{maxryan17}, breaking $\SO(5)$ down to $\O(4)$ makes the situation worse.

In a similar way, one can analyze the three-dimensional stacking of the Shastry-Sutherland lattice. If we allow the possibility of the diamond pVBS phases, the order parameter is written as    $\Phi = (\vect{n}, \Re \scM, \Im \scM)$
where  $\Re \scM$ ($\Im \scM$) represents the square (diamond) pVBS order parameter. Note that now each layer is shifted by $(1,1)$ vector (see \figref{fig:square}) relative to the layer below, and the interlayer AFM coupling is given by \figref{fig:3D_coupling}(a) instead of the direct vertical coupling.  In \figref{fig:3D_coupling}(b), we show two identical layers of N\'eel ordered pattern relatively shifted by $(1,1)$. If we focus on the four spins surrounding the diagonal bond crossing, we found that the two spins from the upper layer and the two spins from the lower layer are aligned oppositely, which is favored by the AFM interlayer spin exchange $J_3$. So the interlayer coupling prefers to lock the N\'eel order parameter in the same direction across the layer as $\vect{n}^{(l)} = \vect{n}^{(l+1)}$. In \figref{fig:3D_coupling}(c), we show two \emph{identical} diamond pVBS patterns displaced from each other. This configuration can gain the effective interlayer ring exchange energy induced by the $J_3$ coupling, which resonates the nearby dimers across the layer. Thus we conclude that $\Re \scM^{(l)}=\Re \scM^{(l+1)}$ is more favorable. In \figref{fig:3D_coupling}(d), we show two \emph{opposite} square pVBS patterns displaced from each other. This configuration also gains interlayer ring exchange energy by resonating the dimers lying on top of each other. But this would require the square pVBS order parameter to be opposite between neighboring layers as $\Im \scM^{(l)}=-\Im \scM^{(l+1)}$. In conclusion, the interlayer coupling prefers $(\vect{n}^{(l)},\Re \scM^{(l)},\Im \scM^{(l)}) = (\vect{n}^{(l+1)},\Re \scM^{(l+1)},-\Im \scM^{(l+1)})$. Here, instead of using $(v_x, v_y)$, we use $(\Re \scM, \Im \scM)$ to parameterize the VBS order parameter, which is a basis change with a positive determinant. In this basis, the interlayer coupling matrix $g$ takes the sign structure of $g^{aa}=(+,+,+,+,-)$. As a result, we again have $\det g <0$, which implies that $\SO(5)$ DQCP would flow to the first order transition point in 3D. Since our $\O(4)$ scenario is considered to be a perturbed $\SO(5)$ DQCP (see \figref{fig:DQCP_metlitski}), it is likely that the $\O(4)$ DQCP would also flow to the first order transition point. }

If $\det g$ happens to be positive, the leading order effect is that the WZW terms would add up together, as the interlayer coupling $g$ tends to lock the order parameters $\Phi^{(l)}$ across the layers. Admittedly, the locking effect will become weaker at longer distance (along the perpendicular direction of layers), but we can still analyze the problem by first grouping the neighboring layers to a renormalized layer and then considering the residual coupling between renormalized layers. Across neighboring layers, the order parameters are expected to bind together, such that the renormalized model can be viewed as a NLSM with WZW term at large level $k$. The intuition from (0+1)D $\O(3)$ WZW term is that the large level limit corresponds to the large spin limit, where the quantum fluctuation of the order parameter is suppressed. Coupling spin-1/2 into a spin chain ferromagnetically in $S_{x,y,z}$-channels results in the ferromagnetic ground state with giant spin and classical spin wave excitations. In this limit, the low-energy physics can be captured within Landau-Ginzberg (LG) theory. Further adding different easy-axis anisotropies to the ferromagnetic spin chain will drive first-order transitions between different Ising ordered phases according to the LG theory.
We conjecture that in higher dimension, similar effect will render each renormalized layer into a classical magnet which should be described within Landau-Ginzberg paradigm, such that the DQCP is not available. So in the presence of interlayer coupling, the N\'eel and VBS phases will likely be separated by a first order transition or intermediate coexisting phases. Thus, our analysis shows that in both $\det g > 0$ and $\det g < 0$ cases, the interlayer coupling would ultimately destabilize the DQCP, and drive it, for example, to a first order transition. However, our analysis also indicates that the $\det g < 0$ case, corresponding to the real material, will have stronger quantum fluctuations, potentially leading to a weaker first order transition or a smaller region of coexisting order parameters than the $\det g > 0$ case. 



One additional remark is that while the interlayer N\'eel order coupling enters directly from the $J_3$ term, the interlayer VBS coupling arises from the higher-order perturbation (resonance) of the $J_3$ term. As a result, the critical point would be shifted to expand the N\'eel order phase. This is consistent with the phase diagram studied in \cite{SS3D_Koga2000}, where the interlayer coupling drives a system into the AFM order and shrink down the VBS region. 




\section{Predictions for Experiment}\label{sec:discussion}

Before discussing experimental consequences of the DQCP, we will make a few remarks on the nature of the plaquette VBS phase. In the Sec.~\ref{sec:sym_analysis}, we discussed two different possibilities for the pVBS phases (see  \figref{fig:pVBS}): the square and diamond pVBS. While the square pVBS breaks the reflection symmetries $\sigma_{xy}$ and $\sigma_{x \bar{y}}$, the diamond pVBS breaks the empty-plaquette-centered $C_4$ rotation symmetry, see \figref{fig:square}. As a result, when the system is at the pVBS phase, magnetic excitations initially degenerate under the $p4g$ symmetry would split differently depending on whether the plaquette is formed at a square or diamond. While our iDMRG simulation of the Shastry-Sutherland model points to the square pVBS phase, in the recent experiments on SrCu$_2$(BO$_3$)$_2$ using INS\cite{Zayed2017} and NMR\cite{Takigawa2007, Takigawa2010No}, the magnetic excitations in the pVBS phase seems to break the $C_4$ rotation symmetry, indicating the diamond pVBS phase. 
This discrepancy implies that the effective spin model for the real material could deviate from the Shastry-Sutherland model studied here. For example, three-dimensional interlayer coupling may induce some effective further-neighbor couplings beyond Shastry-Sutherland model. Therefore the type of pVBS phase to be stabilized at low energy is model dependent. Nevertheless, this does not affect to the  DQCP scenario and the emergent $\O(4)$ symmetry because it only corresponds to a different sign of $\lambda_2$ in \eqnref{eq:LscM}.

As discussed earlier, the DQCP  naturally realizes a quantum spin liquid, a long sought after state of quantum magnets. Furthermore it realizes a particularly exotic variety - a critical spin liquid - with algebraically decaying correlations arising from the gapless emergent degrees of freedom.   Moreover, an experimental realization of the DQCP would be a crucial manifestation of the many-body Berry phase effect that intertwines different order parameters.
A dramatic experimental consequence of the DQCP is the emergent symmetry and resultant spectroscopic signatures expected from INS or RIXS. In particular, the model for SrCu$_2$(BO$_3$)$_2$ studied here  exhibits the $\O(4)$ emergent symmetry with two promising spectroscopic signatures at $X$-point in the Brillouin zone, summarized as the following:
\begin{itemize}[topsep=5pt,parsep=0pt,partopsep=0pt,leftmargin=10pt,labelwidth=6pt,labelsep=4pt]
\item{ Magnon ($S=1$) Channel: This gives the information about the critical fluctuation of the emergent $\O(4)$ conserved current. As a result, the spectral intensity increases linearly with the frequency, $S \sim \omega$. If the emergent symmetry did not exist, there should not exist low energy  spectral weight at this momentum. The deviation from the linear relation would give us a measure of how accurate the emergent symmetry is.}
\item{ Phonon ($S=0$) Channel: This  gives the information about the pVBS order parameter fluctuation. For a given anomalous dimension $\eta_\textrm{VBS}$ of the pVBS order parameter, the spectral intensity diverges with the frequency, $S \sim \frac{1}{\omega^{2-\eta_\textrm{VBS}}}$. The DQCP scenario implies that the VBS order parameter is fractionalized, resulting in a non-zero $\eta_\textrm{VBS}$.   }
\end{itemize}
Moreover, the emergent symmetry implies that $\eta_\textrm{VBS}$ is equal to the anomalous dimension of the N\'eel order parameter fluctuation, $\eta_\textrm{N\'eel}$. However, the N\'eel order parameter fluctuation is located at the $M$-point, which has a pronounced Bragg peak in addition. In principle, the Bragg peak corresponds to spin-0 channel and it must be possible to extract $\eta_\textrm{N\'eel}$ and compare the $\eta_\textrm{N\'eel}$ with $\eta_\textrm{VBS}$ to tell whether the emergent $\O(4)$ symmetry exits. In the earlier work \cite{Ueda2003}, it is estimated from the low-$T$ magnetic susceptibility and heat capacity data that $J_1 \approx 4.7 \textrm{ meV}$, $J_2 \approx 7.3 \textrm{ meV}$, and $J_3 \approx 0.7 \textrm{ meV}$. Moreover, the Debye frequency of the acoustic phonon branch has been measured to be $\omega_D \approx 10 \textrm{ meV}$ in Ref.~\cite{Haravifard2012NSFSDLSSSGS}. Therefore, for experiments to confirm the theoretical predictions, it is required to have the energy resolution smaller than the milli-electron volt. 


According to the present numerical simulation, the phase transition between the pVBS and N\'eel order can be a second-order or weakly-first order transition. The result does not contradict recent numerical work with a similar phase diagram where a first-order transition behavior was observed, because these models \cite{Nahum2018O4, Sandvik2018O4}  are different from the microscopic Hamiltonian in \eqnref{eq:model}, which is more likely to capture the couplings in the real material.
Since the nature of the phase transition may be tunable, it is possible that the experiments on SrCu$_2$(BO$_3$)$_2$ may realize the transition that is either a second order or a weakly first order with  large correlation length. 

Would all these predictions become meaningless if the transition is actually a weakly first order? In fact, even if the two-dimensional system hosts the DQCP, we argued in the previous section that the interlayer coupling might drive the system into a first-order transition point in three-dimension. Indeed, at the weakly first-order transition point, the system would have a finite excitation gap, and experimental spectroscopic data at zero temperature would deviate from \figref{fig:spectrum} due to the absence of a gapless critical fluctuation. However, if we examine the system within the length scale smaller than the (large) correlation length $\xi$, the system would still exhibit the DQCP physics. In other words, if we only examine the system above the energy scale set by the correlation length $\omega >\omega_\text{gap} \sim 1/\xi$, we would observe the predicted spectral intensity trends in \figref{fig:spectrum}. It is our hope that future experiments will be able to use these results to clarify the physics behind the interesting properties of SrCu$_2$(BO$_3$)$_2$.

\section{Conclusions}
We studied a two dimensional $S=1/2$ model which captures key features of the Shastry-Sutherland material SrCu$_2$(BO$_3$)$_2$. We obtained the phase diagram using numerical iDMRG simulations and observed a potentially continuous transition between a plaquette VBS state with two-fold degeneracy and a N\'eel ordered phase.  The transition, studied using both numerical and field theoretical techniques, is proposed to be a deconfined quantum critical point, and we discussed its special features including the lack of a dangerously irrelevant scaling  and an emergent O(4)  symmetry. Concrete predictions are made for future experiments in SrCu$_2$(BO$_3$)$_2$, where a pressure tuned transition between N\'eel order and a putative plaquette VBS state has already been reported \cite{Zayed2017}. The predicted experimental signatures include the form of spectral intensity of spin singlet and spin triplet excitations at extinction points,  which should be accessible in future resonant X-ray and neutron scattering experiments. These can  provide a smoking gun signature of the deconfined criticality and emergent O(4) symmetry. Complications arising from the coupling between layers in the third dimension of the bulk material are briefly discussed, although further work in this direction is needed. We hope this study will trigger future experimental investigation of this  quantum critical point in an interesting material, and more generally provide a road map for the experimental study of deconfined quantum criticality.

\acknowledgements
We thank Y.-C. He, C. Wang, M. Takigawa, A. Sandvik, B. Zhao, L. Wang, Z.-Y. Meng, S.E. Han, and M. Metlitski for stimulating discussions. The DMRG numerics were performed using code developed in collaboration with Roger Mong and the TenPy collaboration. The computations in this paper were run on the Odyssey cluster supported by the FAS Division of Science, Research Computing Group at Harvard University. J.Y. Lee and A. Vishwanath were supported by a Simons Investigator Fellowship. S. Sachdev was supported by the National Science Foundation under Grant No. DMR-1664842.

\appendix

\section{Monopole Transformation}\label{app:monopole}

\begin{figure}[t]
 \hspace{-5pt} \includegraphics[width=0.42\textwidth]{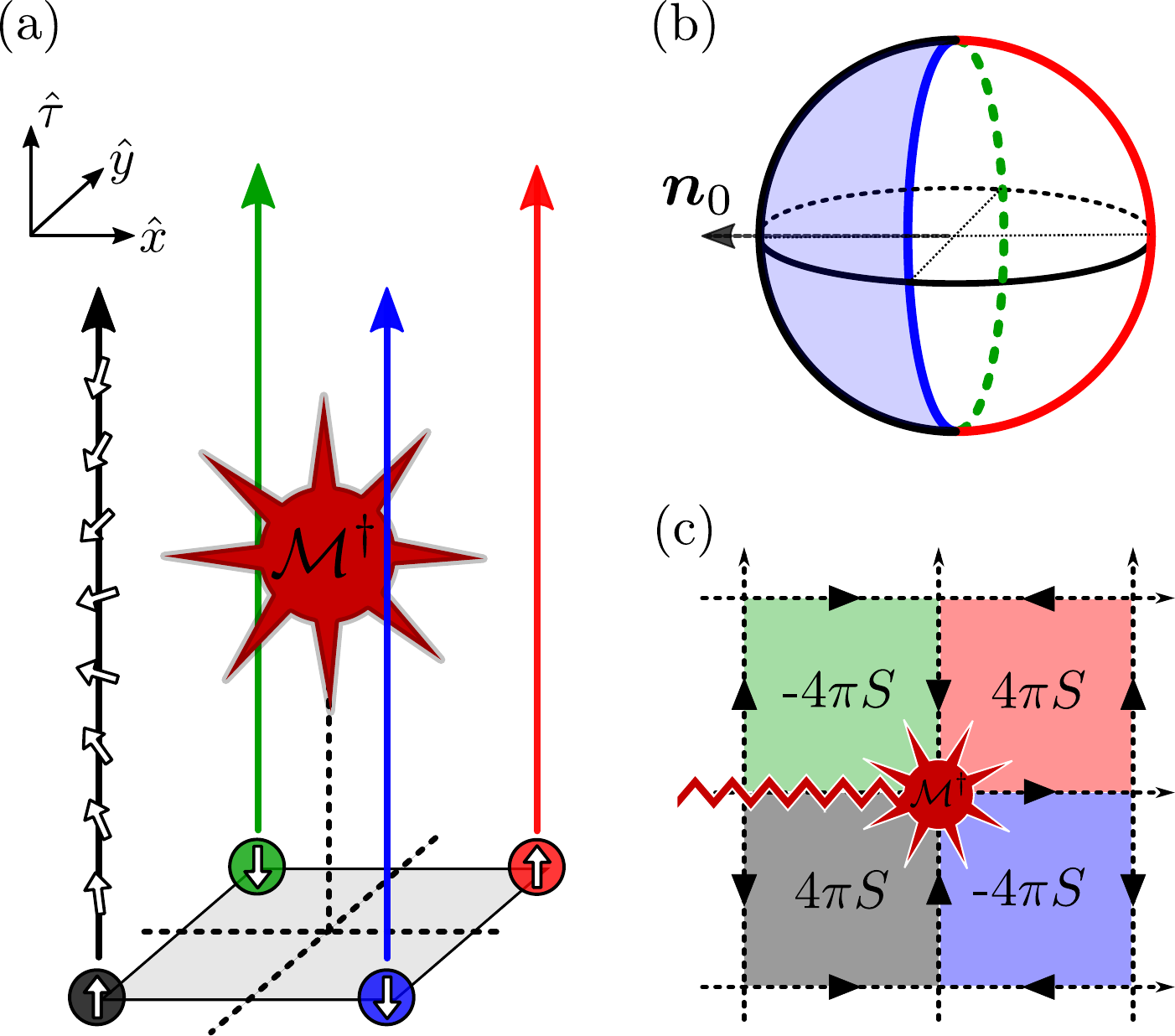}
\caption{ \label{fig:monopole} (a) The monopole operator inserted in the Euclidean spacetime. White arrows represent spin directions. The imaginary time trajectory of each spin is represented by a colored line. White arrows in the black trajectory shows how the direction of the spin changes along the imaginary time when the monopole is inserted. (b) The solid angle $\omega(\vect{n}_r)$ swept through the imaginary time with respect to the reference vector $\vect{n}_0$. (c) The dual lattice where a monopole operator resides on. Here, plaquette centers correspond to the spin sites. The lattice spin acts as an alternating flux pattern $(-1)^{r_x + r_y} 4\pi S$ for monopoles. The hopping amplitude along each black arrow gives a phase factor of $e^{i\pi S} = i$ in our $S=1/2$ case. }
\end{figure}

In this section, we calculate how the monopole operator transforms under the symmetry of the Shastry-Sutherland lattice. To do this, we think of the Shastry-Sutherland lattice as the lattice being deformed from the square lattice with spin-1/2 per site. Starting from the $2+1$D antiferromagnetic ordered phase (N\'eel) of the square lattice, one can derive the action in terms of local N\'eel order parameter $\vect{n}(r_n)$ in a path integral formalism,
\begin{equation}\label{eq:nlsm}
S = \frac{1}{2g} \int d\tau d^2 r \, \qty[ \qty(\frac{\rd \vect{n}}{\rd \tau} )^2 + c^2 (\nabla_r \vect{n})^2 ] + S_B
\end{equation}
where $\vect{n} \sim \epsilon_r \vect{S}$ is a N\'eel order parameter, and $\epsilon_r = (-1)^{r_x + r_y}$ is a factor required for an alternating spin direction. In addition to the continuum action of a classical $O(3)$ nonlinear $\sigma$-model, there exists a Berry phase contribution due to the quantum nature of spin dynamics, which manifestly has the lattice origin. \cite{Haldane1988, Read_Sachdev1990} Let $\omega(n_r)$ be a solid angle swept by a local N\'eel vector located at $r$ throughout the imaginary time evolution from 0 to $\beta$, with respect to the reference direction $n_0$ (See \figref{fig:monopole}(b)). Then, $S_B = i S \sum_{r} \eta_r \omega(\vect{n}_r)$ where $S$ is the spin of an each site and $\eta_r = (-1)^{r_x + r_y}$ is an alternating phase factor coming from the antiferromagnetic nature of the magnetic order.

For any spatial slice, one can define the Skyrmion number $Q(\tau) = \frac{1}{4\pi} \int \hat{\vect{n}} \cdot (\rd_x \hat{\vect{n}} \times \rd_y \hat{\vect{n}} )\Big|_\tau$, which is a topological invariant. Then, a monopole creation operator is defined as a topological defect at the spacetime point which changes $Q(\tau)$ by $+1$ across it. By the further duality mapping, this monopole in the NL$\sigma$M will be mapped into the monopole of the CP$^1$ theory. As the center of the monopole cannot have a finite spin direction, the monopole is located at a dual lattice. When we have a monopole event in the spacetime, it must give a branch-cut structure on the image of $\omega(r)$ because $\omega(r)$ must change by $4\pi$ around the monopole location in the space. More intuitively, when monopole residing on the dual lattice encircles a single site, the imaginary time trajectories of all spins except the one encircled by the monopole would oscillate just back and forth, while the trajectory for the encircled one would entirely wind its $\omega$ by $4\pi$ upon the completion of encircling as in \figref{fig:monopole}(b). Thus, one can view this problem as a monopole hopping around the dual 2D lattice in which each site in the original lattice gives a $4\pi S \eta_r$ flux through the plaquettes of the dual lattice. (Monopole is a charged object under this ``flux'' emanated from a spin-$S$.) The associated phase factor is independent of the exact imaginary time location of the monopole event. Thus for $S=1/2$ case, one can fix the system into a certain gauge and view this as if $\pm \pi/2$ Aharonov-Bohm phase factor gets accumulated for each hopping process for monopoles.

\begin{table}[t]
\begin{center}
\begin{tabular}{ c |  c }
\hline
\hline
\hspace{3pt} Symmetries \hspace{3pt} & \hspace{30pt} Transformations \hspace{30pt} \\
\hline
$T_x$ & ${\cal M}^\dagger \mapsto -i {\cal M}$ \quad $\vect{n} \mapsto -\vect{n}$ \\
$T_y$ & ${\cal M}^\dagger \mapsto i {\cal M}$ \quad $\vect{n} \mapsto -\vect{n}$ \\
$R^\textrm{site}_{\pi/4}$ & ${\cal M}^\dagger \mapsto i {\cal M}^\dagger$ \quad $\vect{n} \mapsto \vect{n}$\\
$R^\textrm{plaq}_{\pi/4}$ & ${\cal M}^\dagger \mapsto -{\cal M}$ \quad $\vect{n} \mapsto -\vect{n}$\\
$\sigma_{x}$ & ${\cal M}^\dagger \mapsto i {\cal M}$ \quad $\vect{n} \mapsto \vect{n}$\\
$\sigma_y$ & ${\cal M}^\dagger \mapsto -i {\cal M}$ \quad $\vect{n} \mapsto \vect{n}$\\
${\cal T}$ & ${\cal M}^\dagger \mapsto {\cal M}$ \quad $\vect{n} \mapsto -\vect{n}$\\
\hline
\hline
\end{tabular}
\end{center}
\caption{Transformation of the monopole operator ${\cal M}$ and N\'eel vector $\vect{n}$ in the field theory of nonlinear sigma model in the N\'eel order phase. }
\label{tab:symmetry2}
\end{table}%

\begin{table}[t]
\begin{center}

\begin{tabular}{ c |  c  | c }
    \hline
    \hline
    \hspace{4pt} $G_{p4g}$ \hspace{4pt} & \hspace{7pt} $G_{p4m}$ \hspace{7pt} & \hspace{20pt} Action \hspace{20pt} \\
    \hline
    $\boldsymbol{T}_x$ & $T_x^2$ & ${\cal M}^\dagger \mapsto {\cal M}^\dagger$ \quad $\vect{n} \mapsto \vect{n}$  \\
    $\boldsymbol{T}_y$ & $T_y^2$ & ${\cal M}^\dagger \mapsto {\cal M}^\dagger$ \quad $\vect{n} \mapsto \vect{n}$  \\
    $\boldsymbol{\sigma}_{xy}$ & $R_{\pi/2} \sigma_x$ & ${\cal M}^\dagger \mapsto {\cal M}$ \quad $\vect{n} \mapsto \vect{n}$  \\ 
    $\boldsymbol{\sigma}_{x\bar{y}}$ & \, $T_x T_y R_{\pi/2} \sigma_y$ \, & ${\cal M}^\dagger \mapsto {\cal M}$ \quad $\vect{n} \mapsto \vect{n}$  \\
    $\boldsymbol{g}_x$ & $T_x \sigma_x$ & \,\,\, ${\cal M}^\dagger \mapsto -{\cal M}^\dagger$ \quad $\vect{n} \mapsto -\vect{n}$   \\
    $\boldsymbol{g}_y$ & $T_y \sigma_y$ & \,\,\, ${\cal M}^\dagger \mapsto -{\cal M}^\dagger$ \quad $\vect{n} \mapsto -\vect{n}$  \\
    $\boldsymbol{R}_{\pi/4}$ & $R^\textrm{plaq}_{\pi/4}$ & ${\cal M}^\dagger \mapsto -{\cal M}$ \quad $\vect{n} \mapsto -\vect{n}$\\
    $ \bf{\cal T} $ & $\cal T$ & ${\cal M}^\dagger \mapsto {\cal M}$ \quad $\vect{n} \mapsto -\vect{n}$  \\
    \hline
    \hline
    \end{tabular}
\end{center}
\caption{The table shows the correspondence between the symmetries of the Shastry-Sutherland lattice ($p4g$) and square lattice ($p4m$). Bold symbols are for symmetries of the Shastry-Sutherland lattice. }
\label{tab:p4g_p4m}
\end{table}

The monopole transformation rule is summarized in \tabref{tab:symmetry2}. $R^\textrm{site}_{\pi/2}$ is a site (spin) centered rotation. Note that it is important to fix the convention for the rotation center because translation symmetry is projective. Here, we choose a rotation to be defined with respect to the black spin in \figref{fig:monopole}(a). $R^\textrm{plaq}_{\pi/2}$ is a plaquette centered rotation, which is defined with respect to the center of four spins in \figref{fig:monopole}(a).  Under the unit translations $T_{x,y}$, time reversal ${\cal T}$, and plaquette centered rotation $R^\textrm{plaq}_{\pi/2}$, the N\'eel order changes its sign. It means that the flux pattern is reversed under such transformations, thus we need to transform a monopole into an anti-monopole to compensate for the change. For reflections $\sigma_{x,y}$, although $\vect{n}$ does not flip, the definition of the Skyrmion number tells that we need to change the sign for the number of monopoles. Thus, a monopole transforms into an anti-monopole again. After figuring out whether a monopole transforms into a monopole or anti-monopole, we need to multiply it by an additional phase factor $\alpha_g$ to account for the Berry phase effect. 

Assume the topological term $S_B$ is absent momentarily, which is how CP$^1$ theory in \eqnref{eq:CP1} is derived. This is a usual practice because unlike the first term inside the parenthesis in \figref{eq:nlsm}, the second term, $S_B$, cannot be straightforwardly extended to the continuum field theory description.  
Under the absence of $S_B$, the monopole insertion operator $\scM^\dagger$ just makes a global adjustment of the N\'eel order configuration to increase the Skyrmion number by one, without any additional phase factor.

However, we know from the existence of $S_B$ in the lattice description that the Berry phase effect is important. 
In order to take into account the Berry phase effect, we need to examine how the monopole transforms under each symmetry action. Under the active transformation where the coordinate system remains the same, we have
\begin{equation}
    g: {\cal M}^\dagger_r \mapsto \alpha_g \cdot g[ {\cal M}^\dagger]_{g (r)}, \quad g[ {\cal M}^\dagger] = {\cal M}^\dagger \text{  or  } {\cal M},
\end{equation}
where the action of $g$ on ${\cal M}^\dagger$ is determined by the previous argument on whether the monopole transforms into the monopole or anti-monopole upon the symmetry transformation.

To determine $\alpha_g$, we need to fix a gauge first. Fixing a gauge is important because a monopole is always created in a pair with an anti-monopole. Thus, the monopole event always has a reference point (anti-monopole) connected by the branch-cut. The Berry phase factor is independent of the way we draw the branch-cut because going around a single spin-lattice site gives a $2\pi$ phase. Let's fix the reference gauge such that the monopole created at $(\bar{0},\bar{0})$ gives a Berry phase $\beta$. Then, for a generic coordinate $r$, inserting a monopole would give the Berry phase factor $ \eta(r) \beta$ where $\eta(r) = 1, i, -1, \textrm{ or }-i$ depending on whether the coordinates $(r_x,r_y)$ are (even, even), (odd, even), (odd, odd), (even, odd) \cite{Haldane1988}. This is shown explicitly in \figref{fig:monopole}(c) as moving the monopole along the arrow gives an additional phase factor $i$. Inserting an anti-monopole at $r$ gives a phase factor $\eta^*(r) \beta^*$ since it would give an exactly opposite contribution to the Berry phase term by having $\omega \mapsto -\omega$. Now, by fixing $\beta = 1$, the insertion of the monopole and anti-monopole at each dual lattice site simply gives a phase $\eta(r)$ and $\eta^*(r)$. 

To illustrate the further procedure, let us consider two examples, $R^\textrm{site}_{\pi/2}$ and $T_x$. 
Under $R^\textrm{site}_{\pi/2}$, a monopole remains monopole, but its location changes as $r \mapsto R^\textrm{site}_{\pi/2} r$. By calculating the relative phase factor between the monopole created at $r$ and $R^\textrm{site}_{\pi/2} r$, we obtain its transformation rule in the continuum description:
\begin{align}
\frac{\eta(R^\textrm{site}_{\pi/2} r) \beta}{\eta(r) \beta} = i \quad \Longrightarrow \quad \scM^\dagger \mapsto  i \scM^\dagger.
\end{align}

In the case of $T_x$, a monopole transforms into an anti-monopole under the symmetry action. At the same time, its location changes as $r \mapsto r+\hat{x}$. Following the similar procedure, we obtain the following rule:
\begin{align}
\frac{ \eta^*(r+\hat{x}) \beta^* }{ \eta(r) \beta} = -i \quad \Longrightarrow \quad \scM^\dagger \mapsto  -i \scM.
\end{align}
Following the similar analysis, we can obtain $\alpha_g$ for all symmetry transformations summarized in \tabref{tab:symmetry2}. In the case of time-reversal symmetry, we do not need an extra phase factor because time-reversal already complex-conjugates the phase factor of a monopole operator, which matches with the phase factor given by the anti-monopole at the same site. Moreover, any monopole condensation amplitude would be time-reversal symmetric because ${\cal T}: \expval{ \scM } \mapsto \expval{ \scM^\dagger }^* = \expval{ \scM}$.

So far, we assumed $\beta=1$. However, a different choice of $\beta$ can be made, which would affect to the transformation rule for the symmetries that filps monopole to anti-monopole. In fact, we can show that this is related to 
the identification rule between monopole ${\cal M}^\dagger$ and the VBS order parameters $v_x$ and $v_y$. For $\beta=1$, we get the relation \eqnref{eq:monopole_VBS}; for example, $T_x:\, \scM^\dagger \mapsto -i \scM$ implies that the $T_x$-invariant monopole condensation corresponds to the condition that $\Re \scM + \Im \scM = 0$. However, if $\beta = \beta=e^{i\pi/4}$, we would get $T_x:\, \scM^\dagger \mapsto -\scM$, implying that the $T_x$-invariant condition is $\Re \scM = 0$. In such a case, we can deduce that $\scM^\dagger \sim v_x + i v_y$.

Once we fix the gauge choice and determine how the monopole transforms under the square lattice symmetries ($p4m$), how the monopole transforms under the Shastry-Sutherland lattice symmetries ($p4g$) can be deduced easily. This is because the Shastry-Sutherland lattice symmetries can be expressed in terms of the square lattice symmetries if we take the Shastry-Sutherland lattice in \figref{fig:square} is deformed from the smaller square lattice. 
The result is summarized in \tabref{tab:p4g_p4m}.


\section{Comparison with a DMRG simulation result of the $J_1$-$J_2$ model in the square lattice}\label{app:numerics}

In this section, we present our analysis on the square lattice with spin-1/2. Using the iDMRG simulation, we studied these models on an infinite cylinder with a circumference size up to $L=10$ lattice sites. Here, we focused on the correlation length spectra. The simulation is explicitly $U(1)_z$ symmetric, and we can plot the correlation spectra for each $U(1)_z$ quantum number, $S_z$. Since following models are $\SO(3)$ symmetric in the microscopic Hamiltonian, there must exist some degeneracy between different $S_z$ sectors, which can be interpreted as the spectrum for a higher spin.

\begin{figure}[t]
\begin{center}
\includegraphics[width=0.45\textwidth]{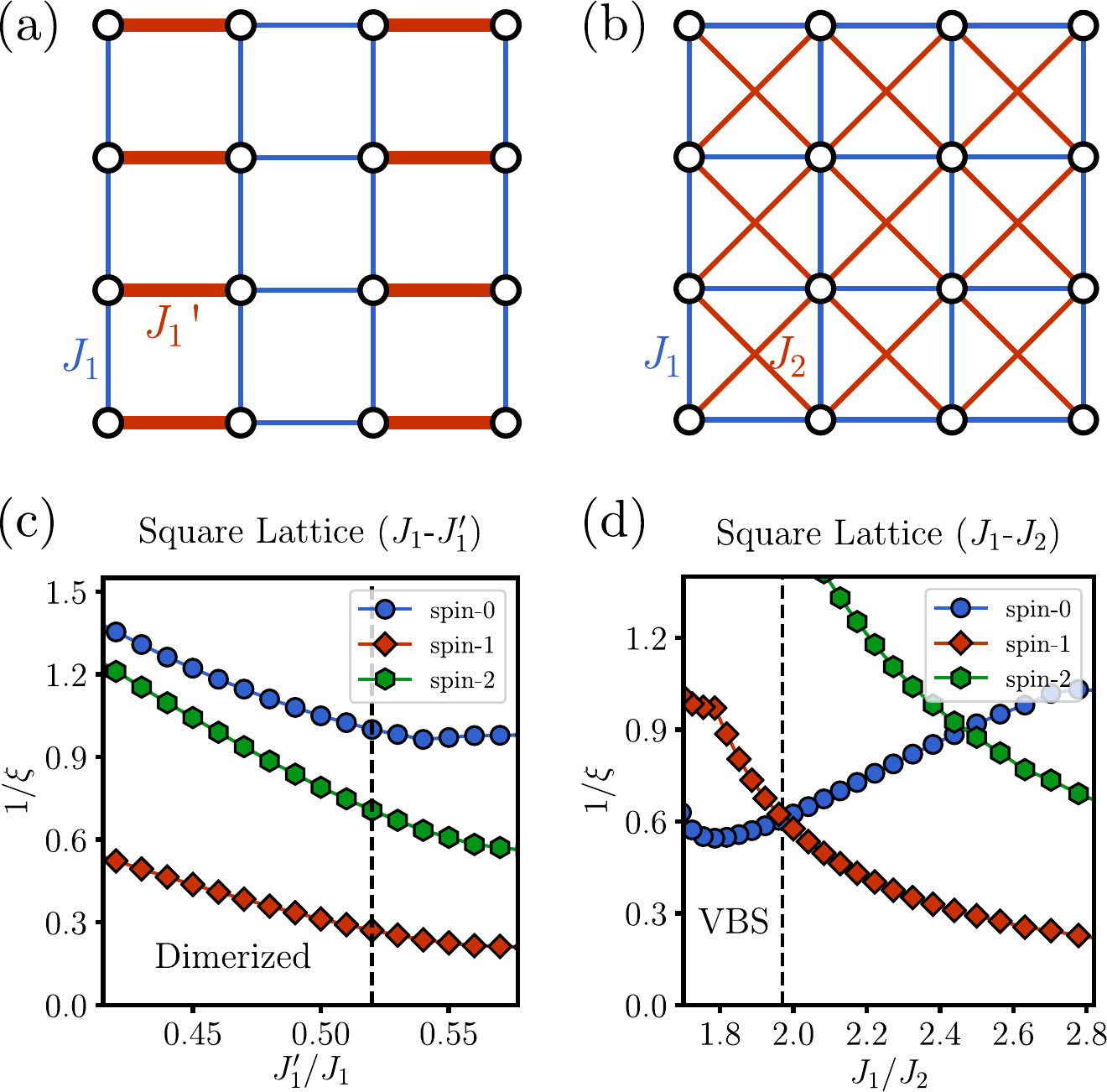}
\caption{(a) Square lattice $J_1$-$J_1'$ model with a dimer coupling $J_1'$. $J_1'$ explicitly breaks the square lattice symmetry. (b) Square lattice $J_1$-$J_2$ model. (c,d) Correlation length spectra for the model in (a,b)}
\label{fig:J1J2_example}
\end{center}
\end{figure}

First, let us consider the case where the square lattice symmetry is broken. The following model realizes the phase transition between N\'eel order and dimerized phase:
\begin{equation}
H = J_1 \sum_{\expval{i,j}\in \textrm{blue}} \vect{S}_i \cdot \vect{S}_j + J'_1 \sum_{ \expval{i,j} \in \textrm{red}} \vect{S}_i \cdot \vect{S}_j 
\end{equation}
where red and blue bonds are shown in \figref{fig:J1J2_example}(a). Here, the dimerized phase does not break any symmetry because the square lattice symmetry is already broken in the model. Therefore, the transition should be described by the Landau-Ginzburg theory. Indeed, it is known from the quantum Monte Carlo simulation \cite{O(3)transition_1} that the system realizes $\O(3)$ Wilson-Fisher critical point at $J_1/J_1' = 0.523$. In the iDMRG simulation, we also observed that the N\'eel order parameter develops at $J_1/J_1' \sim 0.52$. At the transition, a single monopole event is not suppressed because different configurations for single monopole event cannot exactly cancel each other due to the absence of symmetries. As a result, the gauge fluctuation (VBS order parameter fluctuation) becomes confining, and the CP$^1$ theory is no longer valid. Instead, the critical theory is described by the classical NLsM with $\O(3)$ N\'eel vector. The correlation spectra in  \figref{fig:J1J2_example}(c) shows that the spin-triplet correlation length is the largest across the phase transition while the spin-singlet correlation length is much smaller than that. This behavior is consistent with the critical theory described by the classical NLsM.

Next, we study the $J_1$--$J_2$ Heisenberg model with a square lattice symmetry. The model is defined by the following Hamiltonian:
\begin{equation}
H = J_1 \sum_{\expval{i,j}} \vect{S}_i \cdot \vect{S}_j + J_2 \sum_{\expval{\expval{i,j}}} \vect{S}_i \cdot \vect{S}_j 
\end{equation}
where $\vect{S}_i$ is a spin-1/2 operator, $J_1$ is the nearest-neighbor AFM coupling, and $J_2$ is the next nearest-neighbor AFM coupling, see \figref{fig:J1J2_example}(b). When $J_2 = 0$ ($J_1 = 0$), the model is known to realize N\'eel ordered (conventional AFM stripe) phase. For the intermediate value of $J_1/J_2$, the system is frustrated and known to realize the disordered phase.

In accordance with the recent IPEPS study \cite{J1J2_Sheng2018}, we obtained the N\'eel, columnar VBS (cVBS), and conventional stripe phases as we increase $J_2/J_1$. However, in order to obtain the VBS order, we had to apply some bias (pinning field). Under the absence of the bias, the system looks totally symmetric, implying the existence of the symmetric superposition of symmetry broken states, namely a Cat state. The Cat state can be preferred over the symmetry broken phase if the circumference size is comparable to the length scale associated with the fluctuations, which is the size of the monopole in this case.

Before getting into the discussion of the correlation length spectra, we want to elaborate on some simulation details. In our iDMRG simulation, the iDMRG unit cell  consists of two columns of lattices along the circumference, as otherwise translational symmetry broken phases (AFM order or VBS order) cannot develop. The price to pay is that $k_x=0$ and $k_x=\pi$ momenta become indistinguishable. (In our simulation, $k_y$ cannot be measured)
However, at the critical point where the explicit symmetry breaking order has not developed yet, we can use a single column to distinguish $k_x=0$ and $k_x = \pi$ momenta. Indeed, in the iDMRG simulation of the single-column unit cell, we observe that the largest correlation length for $S=1$ spectra carries momentum $k_x = \pi$ in the single-column simulation, which is consistent with the momentum $(\pi,\pi)$ of the gapless magnon in N\'eel order. For $S=0$ case, we obtain that the lowest one carries $k_x = 0$ while the second lowest one carries $k_x = \pi$, which runs almost parallel to the lowest one. They correspond to the $\mathbb{Z}_4$ VBS order parameter fluctuations (spin-singlet) at $(0,\pi)$ and $(\pi,0)$, but the degeneracy is lifted due to the iDMRG geometry which breaks the $C_4$-rotation lattice symmetry.

Surprisingly, in this model, we obtained the  correlation length spectra that exactly agrees with the level crossing behavior of the excitation spectrum in the finite DMRG algorithm (Fig.2 in Ref.~\onlinecite{WangSandvik2018}). Our result is consistent with Ref.~\onlinecite{Zauner2015}, which discussed the agreement between correlation length spectrum and local excitation spectrum in the DMRG simulation. 
Moreover, we want to comment on the argument in Ref.~\onlinecite{WangSandvik2018}. In this previous work, it was argued that the small region where $\xi_{S=1} > \xi_{S=0} > \xi_{S=2}$ corresponds to the gapless spin liquid phase \cite{WangSandvik2018}. However, this reasoning is inconsistent with the iDMRG simulation result of the Shastry-Sutherland lattice because this region is clearly a symmetry broken phase with a non-zero N\'eel order parameter from the numerics. Thus, we can conjecture that such a region  would shrink into a critical `point' rather than remain as an extended phase of Dirac spin liquid both in the $J_1$-$J_2$ square lattice model and Shastry-Sutherland model. Indeed, if we perform a single-column iDMRG simulation for the $J_1$-$J_2$ model, the simulation does not converge well for the $J_1/J_2>2.0$, which means that the hypothesized gapless spin liquid phase is, in fact, more like the AFM phase where the double-column iDMRG unit cell is required.

Finally, we remark on the evidence that supports our discussion in Sec.~\ref{sec:dangerous}. In the previous finite DMRG works on this model \cite{J1J2_Fisher2014}, the plaquette VBS appeared instead of the columnar VBS. In fact, it was found in Ref.~\cite{J1J2_Sheng2018} that these two states have almost the same energy ($\Delta E/E<0.1\%$). This again implies that the dangerously irrelevant operator ${\cal M}^4$, which is responsible for the VBS ordering, has not flowed large enough to condense monopole to a certain direction. This can be supported by Fig.~\ref{fig:J1J2_spectra}(a), where the correlation length for the spin-singlet operator is smaller than the correlation length of the spin-triplet operator throughout the whole intermediate regime between the N\'eel and conventional AFM stripe order.

\section{VBS-Phonon Coupling and Phonon Spectrum}\label{app:phonon}

The square pVBS order breaks the glide reflection symmetries $G_x,G_y$ and the diagonal reflection symmetries $\sigma_{xy}, \sigma_{x\bar{y}}$. Due to the lattice symmetry breaking, the pVBS order should induce lattice distortion as shown in \figref{fig:pVBS}. Therefore the pVBS fluctuation must couple to the lattice vibration mode, i.e. the phonon mode. Here we would like to determine the specific form of the pVBS-phonon coupling.

\begin{figure}[htbp]
\begin{center}
\includegraphics[width=\columnwidth]{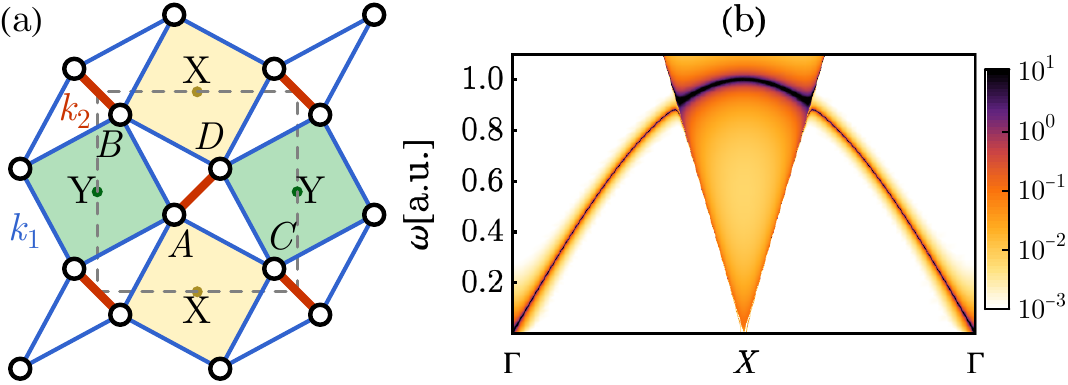}
\caption{(a) $A,B,C,D$ label four copper sites in each unit cell (marked out in dashed lines). $k_1,k_2$ are the stiffnesses of the two types of bonds (nearest neighbor and dimer). $X,Y$ label the two types of plaquettes. (b) Schematic illustration of the phonon spectrum. A continuum will emergent at $X$ point due to the VBS-phonon coupling.}
\label{fig:phonon}
\end{center}
\end{figure}

We will focus on the copper lattice in the following discussion. Although the lattice also contains other atoms and the phonon spectrum can be complicated, we choose to work on the symmetry level to demonstrate the universal consequences of pVBS fluctuation on the phonon spectrum without diving into the details. For this purpose, we first specifies the coordinate of copper sites in the each unit cell. As shown in \figref{fig:phonon}(a), there are four copper sites in each unit cell. At equilibrium, they locate at
\eqs{
\vect{r}_A&=(1+\delta,1+\delta)/2,\\
\vect{r}_B&=(1-\delta,3+\delta)/2,\\
\vect{r}_C&=(3+\delta,1-\delta)/2,\\
\vect{r}_D&=(3-\delta,3-\delta)/2,\\}
where $0\leq \delta<1$ parameterize the deformation of the Shastry-Sutherland lattice from the square lattice. According to Ref.\,\onlinecite{SSmodel1999}, the lattice constant is 8.995\AA, the shortest Cu-Cu bond is 2.905\AA and the second shortest Cu-Cu bond is 5.132\AA. This implies that, in unit of the lattice constant, we have
\eqs{\overline{AD}&=\frac{1+\delta}{2}\approx\frac{2.905\mathrm{\AA}}{8.995\mathrm{\AA}},\\
\overline{AB}&=\frac{\sqrt{1+\delta^2}}{2}\approx\frac{5.132\mathrm{\AA}}{8.995\mathrm{\AA}}.}
The optimal solution is $\delta\approx 0.544$. Using this deformation parameter, we can write down equilibrium positions of copper atoms through out the lattice.

Our strategy to figure out the pVBS-phonon coupling is to first investigate the pattern of lattice distortion induced by the pVBS order. Because the pVBS order does not enlarge the four-copper unit cell, its induced lattice distortion will also be identical among unit cells. Therefore the distortion can be described by four displacement vectors $\vect{u}_{A},\vect{u}_{B},\vect{u}_{C},\vect{u}_{D}$, translating each sublattice separately as
\eq{\vect{r}_i\to\vect{r}'_i=\vect{r}_i+\vect{u}_i\quad(i=A,B,C,D).}
The energy cost associated with the distortion can be modeled by summing up the bond energies
\eqs{\label{eq:Ebond}E_\text{bond}[\vect{u}_i]=&\frac{k_1}{2}\sum_{ij\in\text{n.n.}}\big((\vect{r}'_i-\vect{r}'_{j})^2-(\vect{r}_i-\vect{r}_{j})^2\big)^2\\
+&\frac{k_2}{2}\sum_{ij\in\text{dimer}}\big((\vect{r}'_i-\vect{r}'_{j})^2-(\vect{r}_i-\vect{r}_{j})^2\big)^2,}
where $\vect{u}_i$ dependence is implicit in $\vect{r}'_i=\vect{r}_i+\vect{u}_i$. The energy will increase whenever a bond is stretched or compressed. The shape of the potential in \eqnref{eq:Ebond} captures this physics when the distortion $\vect{u}_i$ is small. The two stiffness coefficients $k_1$ and $k_2$ are expected to be different in general. Of course, in the realistic material, Sr, B, O atoms will all be involved and the bond energy model will be more complicated. However the toy model \eqnref{eq:Ebond} respects all the symmetry property and provides the stiffness to the copper lattice, which can be used to analyze the pVBS induced lattice distortion. Finally we note that the energy model \eqnref{eq:Ebond} written with respect to a single unit cell with periodic boundary condition (i.e. on a torus geometry), so for those bond across the unit cell, their bond length must be correctly treated by considering the periodic boundary condition.

Upon introducing the pVBS order, we will add an additional term to the energy model,
\eqs{\label{eq:E of u}E[\vect{u}_i]&=E_\text{bond}[\vect{u}_i]+E_\text{VBS}[\vect{u}_i],\\
E_\text{VBS}[\vect{u}_i]&=\Im\scM\sum_{p}(-)^p\sum_{i\in p}(\vect{r}'_i-\vect{R}_p)^2,}
where $p$ denotes the square plaquettes and $i\in p$ denotes the four corner sites around the plaquette $p$. $\vect{R}_p$ coordinates the plaquette center,
\eq{\vect{R}_p=\left\{\begin{array}{ll}(1,0) & p\in X,\\(0,1) & p\in Y.\end{array}\right.}
The staggering factor $(-)^p$ is $+1$ for $X$-type plaquette (yellow) and $-1$ for $Y$-type plaquette (green) as shown in \figref{fig:phonon}(a). $\Im\scM=v_y-v_x$ denotes the square pVBS order parameter. The physical meaning of $E_\text{VBS}$ is  that the pVBS order will contract one type of the square plaquette and expand the other type, exerting forces on copper atoms that points towards or away from the plaquette center. 

Given the full energy model in \eqnref{eq:E of u}, we can expand $E[\vect{u}_i]$ to the quadratic order of $\vect{u}_i$. The linear term will be proportional the pVBS order parameter $\Im\scM$, as $\Im\scM$ is the force that distort the lattice. The quadratic term determines how the lattice responses to the distortion force in the linear response regime. We found that independent of the choice of $\delta$ and $k_2$, the response is always given by
\eqs{\label{eq:usol}\vect{u}_A&=\frac{\Im\scM}{4k_1}(-1,1),\\
\vect{u}_B&=\frac{\Im\scM}{4k_1}(-1,-1),\\
\vect{u}_C&=\frac{\Im\scM}{4k_1}(1,1),\\
\vect{u}_D&=\frac{\Im\scM}{4k_1}(1,-1).}
Under the Fourier transformation to the momentum space
\eq{\vect{u}(\vect{q})=\sum_{i}\vect{u}_i e^{\ii\vect{q}\cdot\vect{r}_i},}
the solution in \eqnref{eq:usol} corresponds to
\eq{\label{eq:u=VBS}u_x(\pi,0)\propto-\Im\scM, \quad u_y(0,\pi)\propto\Im\scM.}
This calculation indicates that the square pVBS order will lead to a lattice distortion that corresponds to a the simultaneous condensation of phonon modes $u_x$ at momentum $(\pi,0)$ and $u_y$ at momentum $(0,\pi)$. Given that $\Im\scM=v_y-v_x$, we conclude that there must be a linear coupling between lattice displacement and the VBS order parameter in the form of
\eq{\label{eq:VBS-phonon}\scL_\text{VBS-phonon}=\kappa(v_x u_x+v_y u_y),}
in order to produce the linear response in \eqnref{eq:u=VBS}. The coupling in \eqnref{eq:VBS-phonon} can be further justified by symmetry arguments. \tabref{tab:uvsymm} shows the momentum quantum number and the symmetry transformation of the VBS order parameter $\vect{v}$ and finite-momentum phonon mode $\vect{u}$. One can see $\vect{v} $ and $\vect{u}$ have identical symmetry properties and hence  a linear coupling as in \eqnref{eq:VBS-phonon} is allowed.

\begin{table}[htp]
\begin{center}
\begin{tabular}{c|cccc}
& $v_x$ & $v_y$ & $u_x$ & $u_y$ \\
\hline
$\vect{q}$ & $(\pi,0)$ & $(0,\pi)$ & $(\pi,0)$ & $(0,\pi)$ \\
\hline
$G_x$ & $-v_x$ & $-v_y$ & $-u_x$ & $-u_y$ \\
$G_y$ & $-v_x$ & $-v_y$ & $-u_x$ & $-u_y$ \\
$\sigma_{xy}$ & $v_y$ & $v_x$ & $u_y$ & $u_x$ \\
$\sigma_{x\bar{y}}$ & $v_y$ & $v_x$ & $u_y$ & $u_x$ 
\end{tabular}
\end{center}
\caption{Momentum and symmetry transformations of the VBS order parameter $\vect{v}$ and finite-momentum phonon mode $\vect{u}$.}
\label{tab:uvsymm}
\end{table}%

Given the VBS-phonon coupling, we can investigate the effect of low-energy VBS fluctuation on the phonon spectrum near the DQCP. We first write down the field theory action describing both degrees of freedom,
\eqs{S[\vect{u},\vect{v}]=&\frac{1}{2}\sum_{q}(\omega^2-\Omega_\vect{q}^2)\vect{u}(-q)\cdot\vect{u}(q)\\
-&\frac{1}{2}\sum_{q}G_\text{VBS}^{-1}(q)\vect{v}(-q)\cdot\vect{v}(q)\\
+&\sum_{q}\kappa_{\vect{q}}\vect{v}(-q)\cdot\vect{u}(q),}
where $q=(\omega,\vect{q})$ represents the energy-momentum vector. $\Omega_{\vect{q}}$ describes the phonon dispersion relation. $G_\text{VBS}$ is the correlation function of the VBS critical fluctuation, whose low-energy behavior is given by
\eq{G_\text{VBS}(\omega,\vect{q})=\frac{1}{\big((\vect{q}-\vect{Q})^2-\omega^2\big)^{1-\eta/2}},}
where $\eta$ is the anomalous exponent of the $\O(4)$ vector at the $\O(4)$ DQCP. Based on the previous numerical measurements\cite{YQQin2017,XFZhang2017}, $\eta$ is estimated to be $\eta=0.13\sim0.3$. \change{ $\vect{Q}=(\pi,0)$ or $(0,\pi)$ denotes the momentum point where the VBS fluctuation gets soften. The VBS-phonon coupling $\kappa_\vect{q}$ is expected to be momentum dependent, because the VBS order parameter only couples to the  high-energy phonon around $X$ and $Y$ points but not the acoustic phonon around $\Gamma$ point. By the acoustic phonon around $\Gamma$ point, we mean the low energy part of the phonon, i.e. the ‘segment’ of the acoustic branch around the gapless point, which usually appears in the field theory description of phonons.} Given these setup, we can integrate out the VBS fluctuation and obtain the dressed propagator of phonon,
\eq{D(\omega,\vect{q})=\frac{1}{\Omega_\vect{q}^2-\omega^2-\kappa_\vect{q}^2G_\text{VBS}(\omega,\vect{q})}.}
Then the phonon spectral function can be obtained from
\eq{\label{eq:phononS}S(\omega,\vect{q})=2\Im D(\omega+\ii 0_+,\vect{q}).}
\change{The phonon dispersion $\Omega_\vect{q}$ is unknown to us, as we did not have the full model of the lattice vibration. For demonstration purpose, we can use the following toy model
\eq{\Omega_\vect{q}^2=\sin^2(q_x/2)+\sin^2(q_y/2),}
which captures the gapless acoustic phonon at the $\Gamma$ point and gapped phonons at $X$ and $Y$ points (see \figref{fig:setup}(b)). We also take the anomalous exponent $\eta=0.13$ and use $\kappa_\vect{q}=\kappa_0\Omega_\vect{q}$ with $\kappa_0=0.05$ so that 
\begin{equation}
    \lim_{q \rightarrow (0,\pi),(\pi,0)} \kappa_q = \kappa_0, \qquad     \lim_{q \rightarrow (0,0)} \kappa_q = 0. 
\end{equation} With these, we show the phonon spectrum calculated from \eqnref{eq:phononS} in \figref{fig:phonon}(b).} The prominent feature is a V-shape continuum at the $X$ point (as well as the $Y$ point) in the Brillouin zone. This continuum in the phonon spectrum represents the critical fluctuation of the VBS order parameter at the DQCP. Although the spectral weight is expected to be weak, since the $X$ point is an extinction point, it is still feasible to collect spectral signals of this continuum. In particular, the frequency dependence of the spectral weight at the $X$ point is predicted to follow
\begin{equation}
S(\omega,\vect{q}=X)\propto\omega^{-2+\eta},    
\end{equation}
which can be checked experimentally. It will be meaningful to compare the measured $\eta$ with large-scale quantum Monte Carlo simulation result.

\change{

\section{Detailed Numerical Data}
\label{app:bond_dimension}

\begin{figure}[htbp]
\begin{center}
\includegraphics[width=\columnwidth]{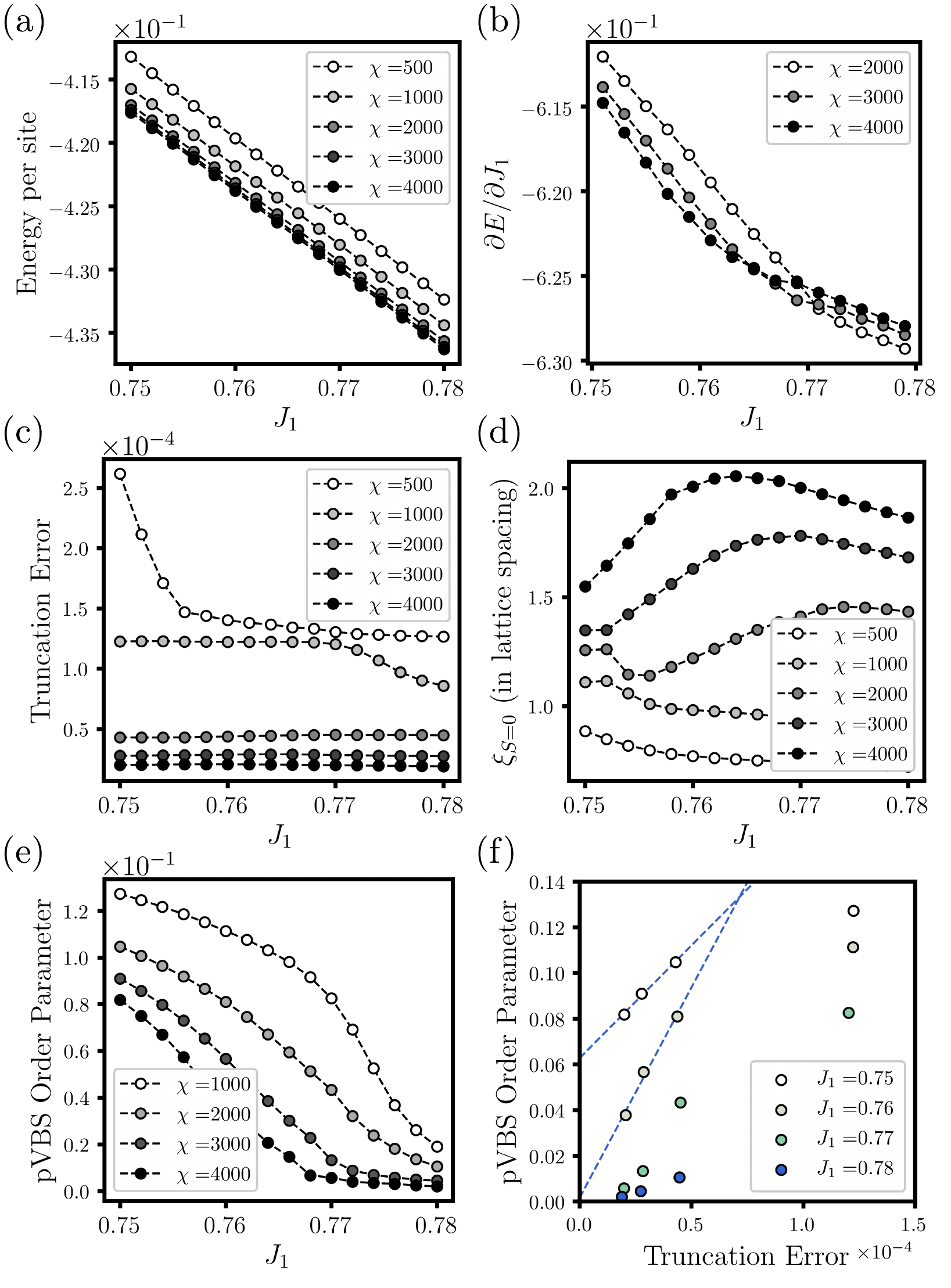}
\caption{ The iDMRG simulation results  with $\Delta J_1 = 0.002$ at $L=10$, shown for a range of MPS bond dimension $\chi$. Bond dimension scalings of (a) energy per site $E$, (b) energy derivative per site $\rd E/\rd J_1$\protect\footnote{Data for $\chi=500,1000$ are not shown here as these data-points behave irregularly, and cover the other data points.}, (c) truncation error $p$, (d) correlation length of spin-singlet operator $\xi_{S=0}$, and (e) pVBS order parameters. Note that the correlation length is plotted instead of the inverse. In (f), we plot pVBS order parameters as functions of truncation erros for a range of the tuning parameter $J_1$. Blue dotted line is a linear fitting for the three data points at $\chi=2000,3000,4000$. }
\label{fig:scaling}
\end{center}
\end{figure}

\begin{figure}[htbp]
\begin{center}
\includegraphics[width=\columnwidth]{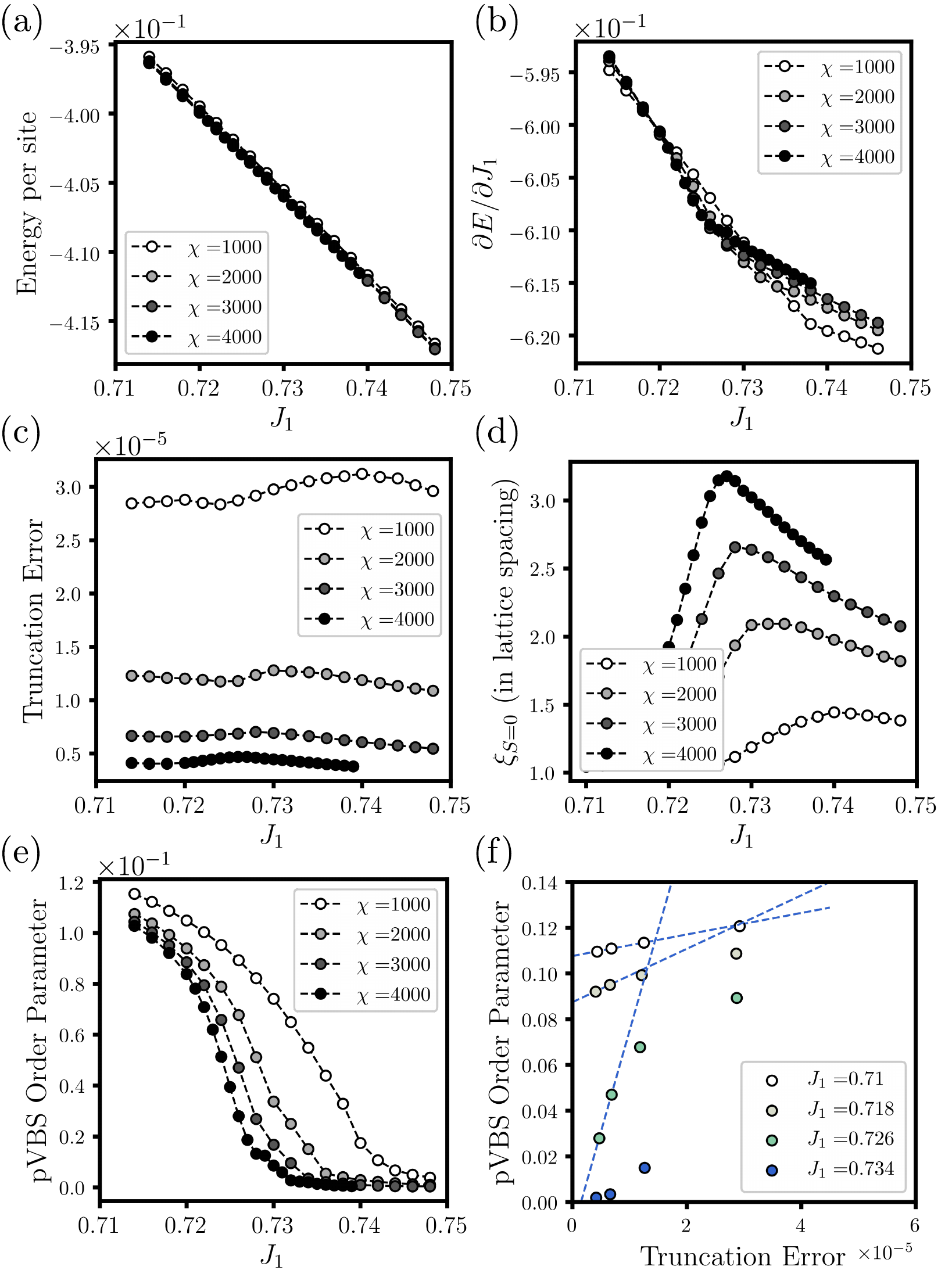}
\caption{ The iDMRG simulation results  with $\Delta J_1 = 0.002$ ($0.001$ for $\chi=4000$) at $L=8$, shown for a range of MPS bond dimension $\chi$. Different observables are labeled as in  \figref{fig:scaling}. Note that the pVBS order parameter vanishes at $J_1 = 0.726$, which is smaller than the value for $L=10$. }
\label{fig:scaling2}
\end{center}
\end{figure}

In this appendix, we discuss the evolution of the iDMRG simulations results as we increase the bond dimension $\chi$ for system sizes $L=8$ and $L=10$. Since the accuracy of the iDMRG simulation is determined by the bond dimension, a reliable analysis requires one to examine the results as a function of the bond dimension. Here, the iDMRG simulation results of the Shastry-Sutherland lattice model in Eq.\ref{eq:model}) at $L=10$ for a range of bond dimensions are presented in Fig.\,\ref{fig:scaling}. Although the the truncation error $\epsilon_\textrm{trun}$ is very large ($>10^{-4}$) at the low bond dimensions, as we increase the bond dimensison upto $\chi=4000$, $\epsilon_\textrm{trun}$ goes below $10^{-5}$ and the iDMRG results becomes sensible.

Fig.\,\ref{fig:scaling}(b) shows the first derivative of energy, whose change of the slope correspsonds to the transition between the pVBS and N\'eel ordererd phases. As the bond dimension increases, the transition point shifts leftward, implying that the parameter regime for the pVBS phase shrinks. The behavior aligns with the intuition that the gapped pVBS phase would be favored over the the gapless N\'eel orderd phase for a low entanglement MPS state. Accordingly, the peak of the spin-singlet correlation length which coincides with the phase transition point also shifts leftward, presented in Fig.\,\ref{fig:scaling}(d). At the same time, the peak of the spin-singlet correlation length  becomes larger and more pronounced as the bond dimension increases, signaling the continuous or weakly first order phase transition.

Although the peak location of the spin-singlet correlation length changes with the bond dimension, a further indication of the phases boundary can be obtained from the order parameter plotted versus the truncation error \cite{trunc_scaling2018, triangular_hubbard2018}. In principle, the ground state is fully symmetric and local order parameter cannot be non-zero. However, in the iDMRG simulation, the numerical process favors a minimally entangled state, giving rise to a non-zero local order parameter in a spontaneous symmetry breaking phase at finite bond dimensions. This can even happen when the system is outside but close to the spontaneous symmetry breaking phase, meaning that one needs to plot the order parameter as a function of the truncation error in order to see whether a non-zero order parameter is truly physical. In Fig.\,\ref{fig:scaling}(f), we plotted the pVBS order parameter defined in Eq.~\ref{eq:order_parameters} as a function of the truncation error, and extrapolated them. We observe that the extrapolated order parameter disappears at $J_1 = 0.76$, which agrees well with the peak location $J_1 = 0.762$ of the spin-singlet correlation length at $\chi=4000$ with $L=10$. As mentioned earlier, this extrapolation method would be benefited a lot if a wider range of bond dimensions is available. However, at $\chi=4000$, each data point already takes about 50 hours of simulations times for 12 multi-threads with 80GB RAM, and both the time and RAM scale roughly as $\chi^2$. Therefore, a significantly higher bond dimension is currently inaccessible at our capacity.

Finally, we remark the scaling of the result for different system sizes. In \figref{fig:scaling2}, we plotted the iDMRG simulation results at $L=8$ for a range of MPS bond dimension as in \figref{fig:scaling}. Note that the truncation error is an order of magnitude smaller than the results at $L=10$. Moreover, the value of $J_1$ where the pVBS order parameter disappears is much smaller for the smaller system size $L$. However, this does not imply that the N\'eel ordered phase develops for $J_1 > 0.726$, as we can see from the correlation length plot in \figref{fig:scaling3}. In \figref{fig:scaling3}(a), while the peak of the spin-singlet correlation length implies that the peak corresponds to the phase transition point for the pVBS phase, the spin-triplet correlation length remains almost constant, which implies that the N\'eel order does not develop, since the N\'eel order would give rise to the increase of the spin-triplet correlation length originated from the gapless magnon excitation. On the other hand, at $L=10$, we observe that the peak of the spin-singlet correlation length is immediately followed by the rise of the spin-triplet correlation length which signals the onset of the N\'eel ordered phase. $L=6$ behavior is similar to that of $L=8$, and the signature of the N\'eel ordered phase, such as the staggering magnetization or the rise of the spin-triplet correlation lengh, is not observed near the pVBS critical point. This is related to the discussion in the main text. For a quasi one-dimensional system with a finite-size circumference size, the spontaneous symmetry breaking of the continuous group would be suppressed due to the disordering effect. As a result, the disappearance of the pVBS ordering is not immediately followed by the N\'eel ordering for a finite circumference system. 


\begin{figure}[htbp]
\begin{center}
\includegraphics[width=\columnwidth]{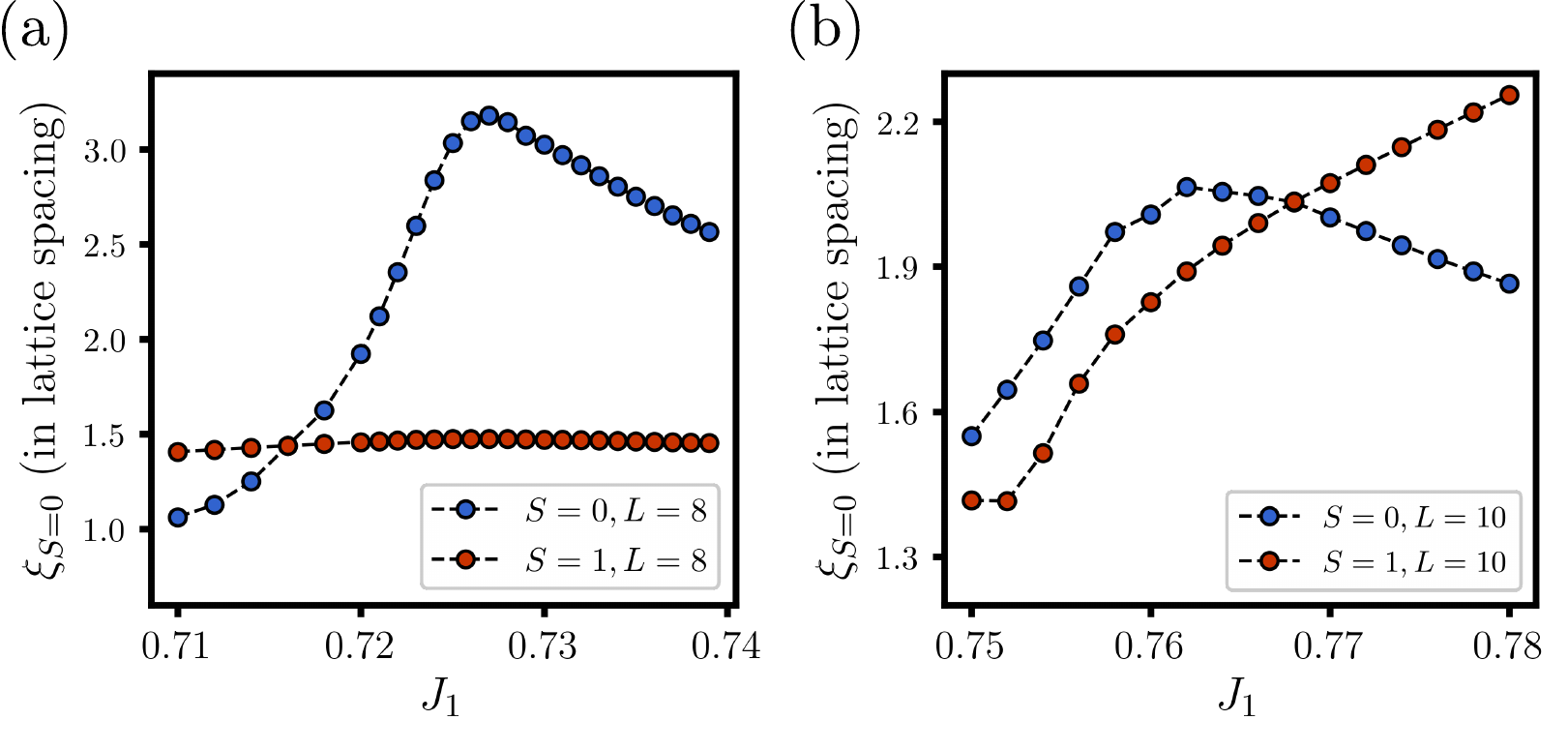}
\caption{ Comparison between the $L=8$ and $L=10$ results at $\chi=4000$ for the spin-singlet and triplet correlation lengths. For $L=8$, the peak of the $\xi_{S=0}$ is located at $J_1 = 0.727$, while for $L=10$ the peak is located at $J_1 = 0.762$. Unlike the result at $L=10$, the spin-triplet correlation length at $L=8$ does not grow much after the peak.  }
\label{fig:scaling3}
\end{center}
\end{figure}

}

\bibliography{DC.bib}

\bibliographystyle{apsrev}
\appendix
\end{document}